\documentstyle[prl,eqsecnum,aps]{revtex}

\begin{document}
\author{Jian-Qi Shen\footnote{E-mail address: jqshen@coer.zju.edu.cn}}
\address{$^{1}$  Centre for Optical
and Electromagnetic Research, State Key Laboratory of Modern
Optical Instrumentation, \\Zhejiang University, Spring Jade,
Hangzhou 310027, P.R. China\\
$^{2}$Zhejiang Institute of Modern Physics and Department of
Physics, Zhejiang University, Hangzhou 310027, P.R. China}
\date{\today }
\title{Novel properties of wave propagation in biaxially anisotropic left-handed materials}
\maketitle

\begin{abstract}
Some physically interesting properties and effects of wave
propagation in biaxially anisotropic left-handed materials are
investigated in this paper. We show that in the biaxially
gyrotropic left-handed material, the left-right coupling of
circularly polarized light arises due to the negative indices in
permittivity and permeability tensors of gyrotropic media. It is
well known that the geometric phases of photons inside a curved
fiber in previous experiments often {\it depend} on the cone
angles of solid angles subtended by a curve traced by the
direction of wave vector of light, at the center of photon
momentum space. Here, however, for the light propagating inside
certain anisotropic left-handed media we will present a different
geometric phase that is {\it independent} of the cone angles. The
extra phases of electromagnetic wave resulting from the
instantaneous helicity inversion at the interfaces between left-
and right- handed (LRH) media is also studied in detail by using
the Lewis-Riesenfeld invariant theory. Some interesting
applications ({\it e.g.}, controllable position-dependent
frequency shift, detection of quantum-vacuum geometric phases and
helicity reversals at the LRH interfaces {\it etc.}) of above
effects and phenomena in left-handed media is briefly discussed.
\\

PACS number(s): 78.20.Ci, 03.65.Vf, 42.50.-p, 42.50.Ct
\end{abstract}
\section{INTRODUCTION}
More recently, a kind of artificial composite metamaterials (the
so-called {\it left-handed media}) having a frequency band where
the effective permittivity and the effective permeability are
simultaneously negative attracts considerable attention of many
authors both experimentally and
theoretically\cite{Smith,Klimov,Shelby,Ziolkowski2,Kong,Garcia,Jianqi}.
In 1967\footnote{Note that, in the literature, some
authors\cite{Kong,Garcia} mentioned the year when Veselago
suggested the {\it left-handed media} by mistake. They claimed
that Veselago proposed or introduced the concept of {\it
left-handed media} in 1968 or 1964. On the contrary, the true
history is as follows: Veselago's excellent paper was first
published in Russian in July, 1967 [Usp. Fiz. Nauk {\bf 92},
517-526 (1967)]. This original paper was translated into English
by W.H. Furry and published again in 1968 in the journal of Sov.
Phys. Usp.\cite{Veselago}. Unfortunately, Furry stated erroneously
in his English translation that the original version of Veselago'
work was first published in 1964.}, Veselago first considered this
peculiar medium and showed from Maxwellian equations that such
media having negative simultaneously negative $\epsilon $ and $\mu
$ exhibit a negative index of refraction, {\it i.e.},
$n=-\sqrt{\epsilon \mu }$\cite{Veselago}. It follows from the
Maxwell's curl equations that the phase velocity of light wave
propagating inside this medium is pointed opposite to the
direction of energy flow, that is, the Poynting vector and wave
vector of electromagnetic wave would be antiparallel, {\it i.e.},
the vector {\bf {k}}, the electric field {\bf {E}} and the
magnetic field {\bf {H}} form a left-handed system; thus Veselago
referred to such materials as ``left-handed'' media, and
correspondingly, the ordinary medium in which {\bf {k}}, {\bf {E}}
and {\bf {H}} form a right-handed system may be termed the
``right-handed'' one. Other authors call this class of materials
``negative-index media (NIM)''\cite{Gerardin}, ``double negative
media (DNM)''\cite{Ziolkowski2} and Veselago's media. It is
readily verified that in such media having both $\varepsilon$ and
$\mu$ negative, there exist a number of peculiar electromagnetic
and optical properties, for instance, many dramatically different
propagation characteristics stem from the sign change of the
optical refractive index and phase velocity, including reversal of
both the Doppler shift and Cherenkov radiation, anomalous
refraction, modified spontaneous emission rates and even reversals
of radiation pressure to radiation tension\cite{Klimov}. In
experiments, this artificial negative electric permittivity media
may be obtained by using the {\it array of long metallic wires}
(ALMWs)\cite{Pendry2}, which simulates the plasma behavior at
microwave frequencies, and the artificial negative magnetic
permeability media may be built up by using small resonant
metallic particles, {\it e.g.}, the {\it split ring resonators}
(SRRs), with very high magnetic
polarizability\cite{Pendry1,Pendry3,Maslovski}. A combination of
the two structures yields a left-handed medium. Recently, Shelby
{\it et al.} reported their first experimental realization of this
artificial composite medium, the permittivity and permeability of
which have negative real parts\cite{Shelby}. One of the potential
applications of negative refractive index materials is to
fabricate the so-called ``superlenses'' (perfect lenses):
specifically, a slab of such materials may has the power to focus
all Fourier components of a 2D image, even those that do not
propagate in a radiative manner\cite{Pendry2000,Hooft}.

In the present paper, we take into consideration the physical
phenomena and effects of circularly polarized photons in biaxially
{\it anisotropic} (gyrotropic) left-handed media. Veselago's
original paper and most of the recent theoretical works discussed
mainly the characteristics of electromagnetic wave propagation
through {\it isotropic} left-handed media, but up to now, the
left-handed media that have been prepared successfully
experimentally are actually {\it anisotropic} in nature, and it
may be very difficult to prepare an isotropic left-handed
medium\cite{Smith,Shelby2,Hu}. Hu and Chui presented a detailed
investigation on the characteristics of electromagnetic wave
propagation in uniaxially anisotropic left-handed media\cite{Hu},
but they concentrated primarily on the classical properties of
wave propagation from the point of view of classical wave optics
and applied electromagnetism. During the last three years, many
researchers in various fields such as materials science, condensed
matter physics, optics and classical applied
electromagnetism\cite{Smith,Klimov,Shelby,Ziolkowski2,Pendry3}
investigated many peculiar optical and electromagnetic properties
in left-handed media. However, to the best of our knowledge, some
physical properties (particularly the purely quantum-mechanical
effects) of polarized photons in left-handed media have not been
considered yet. We think that, in the literature, these problems
get less attention and interest than it deserves. For this reason,
in this paper, these physically interesting properties, phenomena
and effects of wave propagation (polarized photons) in biaxially
anisotropic left-handed materials are investigated in detail.

This paper is organized as follows: in Sec.II, we consider the
left-right (L-R) coupling of circularly polarized light in
biaxially gyrotropic left-handed media; in Sec.III, it is shown
that the geometric phase of photons in a noncoplanarly curved
optical fiber fabricated from biaxially anisotropic left-handed
media is independent of the cone angle of solid angles subtended
by a curve traced by the direction of wave vector of light at the
center of photon momentum space. A scheme of testing
quantum-vacuum geometric phases by using certain biaxially
anisotropic left-handed media is also briefly discussed in this
section; an additional phase acquired by the incident
electromagnetic wave near the interfaces between left- and right-
handed (LRH) media is investigated in detail in Sec.IV, where we
argue that, if, for example, the photons propagating inside a
fiber that is composed periodically of left- and right- handed
(LRH) media, then a new geometric phase of photons due to helicity
inversion will arise. In Sec.V, we conclude with some remarks.
\section{Left-right coupling of circularly polarized light in biaxially gyrotropic left-handed media}
\subsection{Wave propagation in biaxially gyrotropic left-handed media}
It is well known that in ordinary uniaxially gyrotropic media with
the electric permittivity and the magnetic permeability tensors
being of the form
\begin{equation}
(\hat{\epsilon})_{ik}=\left(\begin{array}{cccc}
\epsilon_{1}  & i\epsilon_{2} & 0 \\
-i\epsilon_{2} &   \epsilon_{1} & 0  \\
 0 &  0 &  \epsilon_{3}
 \end{array}
 \right),                 \qquad          (\hat{\mu})_{ik}=\left(\begin{array}{cccc}
\mu_{1}  & i\mu_{2} & 0 \\
-i\mu_{2} &   \mu_{1} & 0  \\
 0 &  0 &  \mu_{3}
 \end{array}
 \right),            \label{eqq200}
\end{equation}
the wave equations of left- and right- handed circularly polarized
light are respectively written

\begin{equation}
\nabla^{2}E_{L}=\frac{n_{L}^{2}}{c^{2}}\frac{\partial^{2}E_{L}}{\partial
t^{2}},  \quad
\nabla^{2}E_{R}=\frac{n_{R}^{2}}{c^{2}}\frac{\partial^{2}E_{R}}{\partial
t^{2}}                          \label{eqq20}
\end{equation}
with the optical refractive indices squared
$n_{L}^{2}=(\epsilon_{1}-\epsilon_{2})(\mu_{1}-\mu_{2})$ and
$n_{R}^{2}=(\epsilon_{1}+\epsilon_{2})(\mu_{1}+\mu_{2})$,
respectively\cite{Veselago}. It is seen that the two circular
modes have different refractive indices. This is referred to as
double circular refraction or circular birefringence. Note that
here both left- and right- handed circularly polarized light are
the eigenmodes of the permittivity and permeability tensors
(\ref{eqq200}). So, $E_{L}$ and $E_{R}$ propagate independently in
the above conventional uniaxially gyrotropic media, {\it i.e.}, no
interaction between them exists in the wave propagation process.

Now we study a completely different case compared with the one
discussed above, namely, it is shown that an interaction between
left- and right- handed circularly polarized light is present in
the wave propagation inside biaxially gyrotropic materials with
left-handed media involved. If, for example, the permittivity and
permeability tensors of the medium considered are written in the
form
\begin{equation}
(\hat{\epsilon})_{ik}=\left(\begin{array}{cccc}
\epsilon_{1}  & i\epsilon_{2} & 0 \\
-i\epsilon_{2} &   -\epsilon_{1} & 0  \\
 0 &  0 &  \epsilon_{3}
 \end{array}
 \right),                 \qquad          (\hat{\mu})_{ik}=\left(\begin{array}{cccc}
-\mu_{1}  & i\mu_{2} & 0 \\
-i\mu_{2} &   \mu_{1} & 0  \\
 0 &  0 &  \mu_{3}
 \end{array}
 \right),              \label{eqq21}
\end{equation}
where $\epsilon_{2}, \mu_{2}$ are real numbers, and
$\epsilon_{1}>0$, $\mu_{1}>0$, then with the help of Maxwell's
equations, one can arrive at
\begin{eqnarray}
\nabla^{2}E_{1}=\left(\frac{\epsilon_{1}\mu_{1}+\epsilon_{2}\mu_{2}}{c^{2}}\right)\frac{\partial^{2}E_{1}}{\partial
t^{2}}-i\left(\frac{\epsilon_{1}\mu_{2}-\epsilon_{2}\mu_{1}}{c^{2}}\right)\frac{\partial^{2}E_{2}}{\partial
t^{2}},                 \nonumber \\
\nabla^{2}E_{2}=\left(\frac{\epsilon_{1}\mu_{1}+\epsilon_{2}\mu_{2}}{c^{2}}\right)\frac{\partial^{2}E_{2}}{\partial
t^{2}}-i\left(\frac{\epsilon_{1}\mu_{2}-\epsilon_{2}\mu_{1}}{c^{2}}\right)\frac{\partial^{2}E_{1}}{\partial
t^{2}}.               \label{eqq22}
\end{eqnarray}
For simplicity, without loss of generality, we assume that the two
mutually perpendicular real unit polarization vectors
${\vec\varepsilon}(k,1)$ and ${\vec\varepsilon}(k,2)$ are taken to
be as follows: $\varepsilon_{1}(k,1)=\varepsilon_{2}(k,2)=1$,
$\varepsilon_{1}(k,2)=\varepsilon_{2}(k,1)=0$ and
$\varepsilon_{3}(k,1)=\varepsilon_{3}(k,2)=0$.  It is readily
verified that the electric field vectors corresponding to left-
and right- handed circularly polarized light are respectively
expressed by $E_{L}=\frac{E_{1}-iE_{2}}{\sqrt{2}}$ and
$E_{R}=\frac{E_{1}+iE_{2}}{\sqrt{2}}$\cite{Bjorken}. It should be
noted that since here $E_{1}$ and $E_{2}$ are complex, $E_{L}$
cannot be considered the complex conjugation of $E_{R}$. From
Eq.(\ref{eqq22}) it follows that the wave equations of left- and
right- handed polarized light propagating inside the biaxially
gyrotropic left-handed materials are obtained as follows
\begin{equation}
\nabla^{2}E_{L}=\frac{n^{2}}{c^{2}}\frac{\partial^{2}E_{L}}{\partial
t^{2}}+\frac{\zeta^{2}}{c^{2}}\frac{\partial^{2}E_{R}}{\partial
t^{2}},   \qquad       \quad
\nabla^{2}E_{R}=\frac{n^{2}}{c^{2}}\frac{\partial^{2}E_{R}}{\partial
t^{2}}-\frac{\zeta^{2}}{c^{2}}\frac{\partial^{2}E_{L}}{\partial
t^{2}},                    \label{eqq25}
\end{equation}
where $n^{2}$ and $\zeta^{2}$ are taken
$n^{2}=\epsilon_{1}\mu_{1}+\epsilon_{2}\mu_{2}$,
$\zeta^{2}=\epsilon_{2}\mu_{1}-\epsilon_{1}\mu_{2}$. Thus the
above equations show that the coupling of left-handed light to the
right-handed one arises from both the {\it gyrotropic} and {\it
left-handed} properties of materials. This interaction can be
treated by introducing two frequency shifts ($\Omega_{L}$ and
$\Omega_{R}$) to the electromagnetic wave amplitudes, namely,
$E_{L}$ and $E_{R}$ can be written in the following
form\cite{Shene-print}
\begin{equation}
 E_{L}\sim \exp
\left\{\frac{1}{i}[(\omega+\Omega_{L})t-\frac{n\omega}{c}z]\right\},
\quad
 E_{R}\sim \exp
\left\{\frac{1}{i}[(\omega+\Omega_{R})t-\frac{n\omega}{c}z]\right\},
\label{eqq26}
\end{equation}
where it has been assumed that the waves propagate along the
$\hat{{\rm z}}$-direction in Cartesian coordinate system.

It is worth noticing that, differing from the double circular
refraction expressed in (\ref{eqq20}), where $E_{L}$ and $E_{R}$
have the same frequency but different optical refractive indices,
here, on the contrary, the two modes in (\ref{eqq26}) have the
same refractive index but different frequencies. The latter case
may therefore be considered a {\it time} analogue to the former
one.

 Substitution of these two expressions (\ref{eqq26}) into the wave
equations (\ref{eqq25}) yields
\begin{equation}
n^{4}(2\omega+\Omega_{L})(2\omega+\Omega_{R})\Omega_{L}\Omega_{R}+\zeta^{4}(\omega+\Omega_{L})^{2}(\omega+\Omega_{R})^{2}=0,
\label{eqq27}
\end{equation}
which is a restriction imposed on the two frequency shifts
$\Omega_{L}$ and $\Omega_{R}$.

It should be noted that the above-presented treatment of wave
propagation in generalized gyrotropic media may be applicable to
the uniaxially anisotropic left-handed materials and this kind of
media will also give rise to the left-right coupling of circularly
polarized light. The permittivity and permeability tensors of
uniaxially gyrotropic left-handed media, which can be actually
fabricated by current technology\cite{Hu}, are written

\begin{equation}
(\hat{\epsilon})_{ik}=\left(\begin{array}{cccc}
\epsilon  & i\epsilon_{2} & 0 \\
-i\epsilon_{2} &   -\epsilon' & 0  \\
 0 &  0 &  \epsilon
 \end{array}
 \right),                 \qquad          (\hat{\mu})_{ik}=\left(\begin{array}{cccc}
-\mu'  & i\mu_{2} & 0 \\
-i\mu_{2} &   \mu & 0  \\
 0 &  0 &  \mu
 \end{array}
 \right),              \label{eqq210}
\end{equation}
where the parameters ({\it i.e.}, $\epsilon$, $\epsilon'$, $\mu$
and $\mu'$) in the tensors are all positive. It is apparently seen
that the L-R coupling of polarized light is present in this type
of materials.

Additionally, note that in both biaxially {\it gyroelectric} (with
$\mu_{2}$=0) and {\it gyromagnetic} (with $\epsilon_{2}$=0)
left-handed media, the above left-right coupling of polarized
light will also occur.
\subsection{Discussions: nonlocal effects and potential applications}
What is the physical origin of left-right coupling of polarized
light in biaxially gyrotropic left-handed media? Here we offer an
interpretation as to why this effect might occur via nonlocal
polarization effect and left-handed properties (negative index) of
media: specifically, neither left- nor right- handed circularly
polarized light is the eigenmode of the permittivity and
permeability tensors (\ref{eqq21}). For this reason, the
interaction between left- and right- handed circularly polarized
light may arise in this type of media.

The electric-dipole approximation method shows that if the optical
wavelength $\lambda$ is large compared with the dimension $\iota$
of the polarizable units ({\it e.g.}, electric dipoles), then the
induced polarization at a point will only be a spatially local
function of the electric field at the same point. This phenomenon
often appears in isotropic media, where certain optical effects
associated with the small ratio $\frac{\iota}{\lambda}$ can be
ignored. In anisotropic media, however, the spatial variations of
the fields over the polarizable units ({\it i.e.}, the magnetic
dipoles and electric quadrupoles that are induced by the applied
external fields) should also be taken into account. This
generalized polarization, the physical quantity of which is no
longer denoted by a local function of the electromagnetic fields,
will also oscillate and emit radiation which adds to the
contribution from the oscillating electric dipoles. It is known
that the spatial nonlocality (spatial dispersion) of generalized
polarization in electromagnetic media leads to the natural optical
activity (natural gyrotropy), which results from the lowest-higher
nonlocal terms in the expansion series of induced polarization,
and some electro-optical effects and magnetic-field induced
gyrotropies, which arises from the terms with a higher-order
nonlocality\cite{PU}. In the electro-optical effects ({\it e.g.},
the Pockels effect and Kerr effect) and magnetic-field induced
gyrotropy ({\it i.e.}, the Faraday rotation effect), the
anisotropism of permittivity and permeability is often considered
the nonlinear optical phenomenon (nonlinear polarization effect),
since the electric and magnetic field strengths occur more than
once in the nonlocal terms of expansion of induced polarization.

We think that the L-R coupling of polarized light can be regarded
as the nonlocal optical effect in character. But this coupling
does not appear in regular gyrotropic media, the electromagnetic
response of which is described by (\ref{eqq200}), where both the
left- and the right- handed circularly polarized light are the
eigenmodes of the permittivity and permeability tensors and no
interaction between these eigenmodes occurs. Only the gyrotropic
anisotropism of permittivity and permeability is combined with
left-handed features of media (where neither the left- nor the
right- handed circularly polarized light is the eigenmode of the
permittivity and permeability tensors) will the L-R coupling of
polarized light be realized. In brief, the physical origin of L-R
coupling of polarized light lies in the nonlocal polarization and
left-handed properties of media.

Generally speaking, it follows from the Kramers-Kronig relation
that the permittivity and permeability of composite materials are
often both dispersive and absorptive. If the imaginary parts of
optical constants $\epsilon_{1,2}$ and $\mu_{1,2}$ in
(\ref{eqq21}) are not small in comparison with their real parts,
then $\zeta^{2}$ is a complex number and consequently both
$\Omega_{L}$ and $\Omega_{R}$ are complex also. This, therefore,
means that the electromagnetic wave amplitudes $E_{L}$ and $E_{R}$
will exponentially decrease in the time evolution process, rather
than only along the path.

In what follows we consider a potential application of L-R
coupling to controlling the behavior of lightwave, {\it i.e.}, the
so-called {\it controllable position-dependent frequency shift} in
certain inhomogeneous media, {\it e.g.}, photonic crystals
composed of such anisotropic left-handed media. Photonic crystals
are artificial materials patterned with a periodicity in
dielectric constant, which can create a range of forbidden
frequencies called a photonic band gap\cite{Yablonovitch,Li}. Such
dielectric structure of crystals offers the possibility of molding
the flow of light. It follows from (\ref{eqq25}) that if both
$\frac{\partial n}{\partial z}$ and $\frac{\partial
\zeta}{\partial z}$ are negligibly small, then the frequency
shifts $\Omega_{L}$ and $\Omega_{R}$ of LRH polarized light due to
the left-right coupling depend upon the spatial position in
photonic crystals. This enables us to control the frequencies of
circularly polarized light by making use of the spatial structures
(or distribution) of optical refractive indices in crystals.

In addition, it is believed that the influence of L-R coupling of
polarized light on the photonic band gap structure is of physical
interest and therefore deserves investigation, since both the wave
vector and the frequency of light are dependent on the spatial
positions in this kind of biaxially gyrotropic left-handed
photonic crystals.

\section{Geometric phases independent of cone angle}
Since Berry discovered that a topological (geometric) phase exists
in quantum mechanical wave function of time-dependent
systems\cite{Berry}, geometric phase problems have captured
intensive attention of researchers in a variety of fields such as
quantum mechanics\cite{AA}, differential geometry\cite{Simon},
gravity theory\cite {Furtado,Shen2}, atomic and molecular
physics\cite {Kuppermann,Kuppermann2,Levi}, nuclear physics\cite
{Wagh}, quantum optics\cite{Gong}, condensed matter
physics\cite{Taguchi,Falci}, molecular structures and molecular
chemical reaction\cite{Kuppermann} as well. More recently, many
authors concentrated on their special attention on the potential
applications of geometric phases to the geometric (topological)
quantum computation, quantum decoherence and related
topics\cite{Wangzd,Wangxb}. One of the most interesting
realizations of Berry's phase ({\it i.e.}, cyclic adiabatic
geometric phase) is the propagation of photons inside a helically
curved optical fiber, which was first proposed by Chiao and
Wu\cite{Chiao}, and performed experimentally by Tomita and
Chiao\cite{Tomita}. Afterwards, a large number of investigators
treated this geometric phase problem by making use of the
classical Maxwell's electrodynamics, differential geometry method
(parallel transport) and quantum adiabatic theory\cite{Berry} both
theoretically and
experimentally\cite{Kwiat,Ross,Robinson,Dresden,Haldane1,Haldane2,Chiao2,Zhou}.
Based on the above investigations, we studied the nonadiabatic
noncyclic geometric phases of photons propagating inside a
noncoplanarly curved optical fiber by using the Lewis-Riesenfeld
invariant theory and the invariant-related unitary transformation
formulation\cite{Lewis,Gao,Gao2}. In the published
paper\cite{Shen1}, we considered the photon helicity inversion in
the curved fiber and its potential applications to information
science and, on the basis of the second-quantized spin model, we
calculated the {\it quantum-vacuum geometric phases} of
electromagnetic fields, which results from the vacuum zero-point
fluctuation.
 \subsection{Nonadiabatic noncyclic geometric phases of photons in the noncoplanar optical fiber}
In what follows we consider a potential vacuum effect in a
time-dependent quantum system, {\it i.e.}, time evolution of
photon wave function and rotations of polarization planes in a
noncoplanar fiber. The spin angular momentum operators of photon
fields read (in the natural units $\hbar=c=1$)
\begin{equation}
S_{ij}=-\int{(\dot{A}_{i}A_{j}-\dot{A}_{j}A_{i})}{\rm d}^{3}{\bf
x} \label{eq21}
\end{equation}
with the three-dimensional magnetic vector potentials ${\bf
A}({\bf x}, t)$ being expanded as a Fourier
series\cite{Bjorken,Lurie}
\begin{equation}
{\bf A}({\bf x},t)=\frac{1}{\sqrt{V}}\sum_{\bf
k}\frac{1}{\sqrt{2\omega_{\bf k}}}\sum_{\lambda=1}^{2}{\vec
\varepsilon}(k,\lambda)[a(k,\lambda)\exp(-i{k\cdot
x})+a^{\dagger}(k,\lambda)\exp(i{k\cdot x})],          \label{32}
\end{equation}
where the frequencies $\omega_{\bf k }=|{\bf k }|$, and ${\vec
\varepsilon}(k,1)$ and ${\vec \varepsilon}(k,2)$ are the two
mutually perpendicular real unit polarization vectors, which are
also orthogonal to the wave vector ${\bf k}$ of the time harmonic
electromagnetic wave.

Consider a noncoplanarly curved optical fiber that is wound
smoothly on a large enough diameter\cite{Tomita}, the effective
Hamiltonian that describes the time evolution of photon
wavefunction in the curved fiber is\cite{Shen1}
\begin{equation}
H_{\rm eff}(t)=\frac{{\bf{k}}(t)\times
\dot{\bf{k}}(t)}{k^{2}}\cdot \bf{S}        \label{eq220}
\end{equation}
with the wave vector being defined to be  ${\bf
k(t)}=k(\sin\lambda \cos\gamma, \sin\lambda  \sin \gamma ,
\cos\lambda)$, where $\dot{\bf{k}}(t)$ denotes the derivative of
${\bf{k}}(t)$ with respect to time $t$. Note that here the wave
vector ${\bf k(t)}$ of a photon propagating inside the fiber is
always along the tangent to the curved fiber at each point at
arbitrary time. Readers may be referred to the Appendices to this
paper for the derivation of effective Hamiltonian (\ref{eq220}).
According to the Lewis-Riesenfeld invariant theory\cite{Lewis} and
the invariant-related unitary transformation
formulation\cite{Gao}, the exact particular solution to the
time-dependent Schr\"{o}dinger equation
\begin{equation}
i\frac{\partial \left| \sigma ,{\bf{k}}(t)\right\rangle }{\partial t}=\frac{%
{\bf{k}}(t)\times \dot{\bf{k}}(t)}{k^{2}}\cdot {\bf{S}}\left|
\sigma ,{\bf{k}}(t)\right\rangle         \label{eq221}
\end{equation}
governing the propagation of photons in the fiber is given by
\begin{equation}
\left| \sigma ,{\bf{k}}(t)\right\rangle=\exp \left[\frac{1}{i}\phi
_{\sigma }^{\rm (g)}(t)\right]V(t)\left| \sigma,k \right\rangle,
\label{eq222}
\end{equation}
where $\left| \sigma,k \right\rangle\equiv\left|
\sigma,{\bf{k}}(t=0) \right\rangle$ is the initial photon
polarized state, and $V(t)=\exp
[\beta (t)S_{+}-\beta ^{\ast }(t)S_{-}]$ \cite{Gao,Shen3} with the time-dependent parameters $\beta (t)=-\frac{\lambda (t)}{2}\exp [-i\gamma (t)]$, $\beta ^{\ast }(t)=-\frac{%
\lambda (t)}{2}\exp [i\gamma (t)]$. The geometric phase of photons
whose initial helicity eigenvalue is $\sigma $ can be expressed
by\cite{Shen1,Zhu,Gaoxc}
\begin{equation}
\phi _{\sigma }^{\rm
(g)}(t)=\left\{{\int_{0}^{t}\dot{\gamma}(t^{^{\prime
}})\left[1-\cos \lambda (t^{^{\prime }})\right]{\rm d}t^{^{\prime
}}}\right\}\left\langle \sigma,k \right| S_{3}\left| \sigma,k
\right\rangle .            \label{eq224}
\end{equation}

In the adiabatic process where both the precessional frequency
$\dot{\gamma}$ (expressed by $\Omega$) and $\lambda$ (${\bf k }$
deviating from the third axis in the fixed frame by an angle
$\lambda$) can be regarded as constants (which can be realized in
a helically coiled fiber\cite{Chiao,Tomita}), the adiabatic
geometric phase ({\it i.e.}, Berry's topological phase) in a cycle
($T=\frac{2\pi}{\Omega}$) in the photon momentum ${\bf k}$ space
is written
\begin{equation}
\phi _{\sigma }^{\rm (g)}(T)=2\pi(1-\cos \lambda)\left\langle
\sigma,k \right| S_{3}\left| \sigma,k \right\rangle,
                                                        \label{eq225}
\end{equation}
where $2\pi(1-\cos \lambda)$ is equal to a solid angle subtended
at the origin of momentum ${\bf k}$ space. This fact thus means
that the geometric phase (\ref{eq224}) or (\ref{eq225}) carries
information on the global and topological properties of time
evolution of quantum systems. It should be emphasized that here
the {\it quantum-vacuum geometric phase} may be involved in
$\left\langle \sigma,k \right| S_{3}\left| \sigma,k \right\rangle$
if the third component $S_{3}$ of photon spin operator ${\bf S}$
is of a non-normal-order form, which will be taken into account in
the following.

Substituting the Fourier expansion series (\ref{32}) of {\bf
A({\bf x},t)} into the expression (\ref{eq21}) for photon spin
operator, one can obtain the non-normal-order photon ${\bf S}$
\cite{Shen1,Zhu}, {\it i.e.},
\begin{equation}
S_{3}=\frac{i}{2}[a(k,1)a^{\dagger}(k,2)-a^{\dagger}(k,1)a(k,2)-a(k,2)a^{\dagger}(k,1)+a^{\dagger}(k,2)a(k,1)].
\label{eq31}
\end{equation}
In what follows we define the creation and annihilation operators,
$a_{R}^{\dagger}(k)$, $a_{L}^{\dagger}(k)$, $a_{R}(k)$,
$a_{L}(k)$, of right- and left- handed circularly polarized light
\cite{Bjorken}
\begin{eqnarray}
a_{R}^{\dagger}(k)&=&\frac{1}{\sqrt{2}}[a^{\dagger}(k,1)+ia^{\dagger}(k,2)],
\quad
a_{R}(k)=\frac{1}{\sqrt{2}}[a(k,1)-ia(k,2)],           \nonumber \\
a_{L}^{\dagger}(k)&=&\frac{1}{\sqrt{2}}[a^{\dagger}(k,1)-ia^{\dagger}(k,2)],
\quad a_{L}(k)=\frac{1}{\sqrt{2}}[a(k,1)+ia(k,2)]. \label{eq32}
\end{eqnarray}
So, the third component of monomode-photon spin operator can be
rewritten
\begin{equation}
S_{3}=\frac{1}{2}\left\{{\left[a_{R}(k)a_{R}^{\dagger}(k)+a_{R}^{\dagger}(k)a_{R}(k)]-[a_{L}(k)a_{L}^{\dagger}(k)+a_{L}^{\dagger}(k)a_{L}(k)\right]}\right\}.
\label{eq39}
\end{equation}
The monomode multi-photon states of left- and right- handed (LRH)
circularly polarized light (at $t=0$) can be defined
\begin{equation}
|\sigma=-1,k,
n_{L}\rangle=\frac{[a_{L}^{\dagger}(k)]^{n}}{\sqrt{n!}}|0_{L}\rangle,
\quad |\sigma=+1,k,
n_{R}\rangle=\frac{[a_{R}^{\dagger}(k)]^{n}}{\sqrt{n!}}|0_{R}\rangle
\label{eq310}
\end{equation}
with $n_{L}$ and $n_{R}$ being the LRH polarized photon occupation
numbers, respectively. Now we calculate the geometric phases of
multi-photon states
\begin{equation}
|\sigma=+1,k, n_{R}; \sigma=-1,k, n_{L}\rangle\equiv|\sigma=+1,k,
n_{R}\rangle\otimes|\sigma=-1,k, n_{L}\rangle \label{eq3100}
\end{equation}
in the fiber. Substitution of (\ref{eq3100}) into (\ref{eq224})
yields
\begin{equation}
\phi^{\rm (g)}(t)=\left\{{\int_{0}^{t}\dot{\gamma}(t^{^{\prime
}})\left[1-\cos \lambda (t^{^{\prime }})\right]{\rm d}t^{^{\prime
}}}\right\}\left\langle \sigma=+1,k, n_{R}; \sigma=-1,k, n_{L}
\right| S_{3}|\sigma=+1,k, n_{R}; \sigma=-1,k, n_{L}\rangle .
\label{eq311}
\end{equation}
and the final result is given
\begin{equation}
 \phi^{\rm (g)}(t)=(n_{R}-n_{L})\left\{{\int_{0}^{t}\dot{\gamma}(t^{^{\prime
}})\left[1-\cos \lambda (t^{^{\prime }})\right]{\rm d}t^{^{\prime
}}}\right\},                                 \label{eq312}
\end{equation}
which is independent of $k$ (the magnitude of ${\bf k}$) but
dependent on the geometric nature of the pathway (expressed in
terms of $\lambda$ and $\gamma$) along which the light wave
propagates. This fact indicates that geometric phases possesses
the topological and global properties of time evolution of quantum
systems. It is emphasized that the phases (\ref{eq312}) associated
with the photonic occupation numbers $n_{R}$ and $n_{L}$ are
quantal in character\cite{Gao}. Gao has shown why $\phi^{\rm
(g)}(t)=(n_{R}-n_{L})\{{\int_{0}^{t}\dot{\gamma}(t^{^{\prime
}})\left[1-\cos \lambda (t^{^{\prime }})\right]{\rm d}t^{^{\prime
}}}\}$ is referred to as the quantal geometric phases\cite{Gao} by
taking into consideration the uncertainty relation between the
operators $\frac{1}{2}[a_{R(L)}^{\dagger}(k)+a_{R(L)}(k)]$ and
$\frac{i}{2}[a_{R(L)}^{\dagger}(k)-a_{R(L)}(k)]$. Although the
phases $\phi^{\rm (g)}(t)$ in (\ref{eq312}) are quantal geometric
phases of photons, they do not belong to the geometric phases at
quantum-vacuum level which arise, however, from the zero-point
electromagnetic energy of quantum vacuum fluctuation. Since the
cyclic adiabatic cases of (\ref{eq312}) have been measured
experimentally by Tomita and Chiao {\it et
al.}\cite{Tomita,Kwiat,Ross,Haldane1}, we will not consider them
further. In the following we will study instead the geometric
phases at quantum vacuum level, which has not been tested
experimentally yet. The reason for why it cannot be easily tested
will be given at the end of this section.

According to the expression (\ref{eq39}) for $S_{3}$, both the
geometric phases of left- and right- handed circularly polarized
photon states, {\it i.e.}, $|\sigma=-1,k, n_{L}\rangle$ and
$|\sigma=+1,k, n_{R}\rangle$, are respectively of the form
\begin{equation}
 \phi_{L}^{\rm (g)}(t)=-(n_{L}+\frac{1}{2})\left\{{\int_{0}^{t}\dot{\gamma}(t^{^{\prime
}})\left[1-\cos \lambda (t^{^{\prime }})\right]{\rm d}t^{^{\prime
}}}\right\},                \quad
 \phi_{R}^{\rm (g)}(t)=+(n_{R}+\frac{1}{2})\left\{{\int_{0}^{t}\dot{\gamma}(t^{^{\prime
}})\left[1-\cos \lambda (t^{^{\prime }})\right]{\rm d}t^{^{\prime
}}}\right\}.                 \label{eq313}
\end{equation}
It follows that the time-dependent zero-point energy possesses
physical meanings and therefore contributes to geometric phases of
photon fields. Thus the noncyclic nonadiabatic geometric phases of
left- and right- handed polarized states at quantum-vacuum level
are given
\begin{equation}
 \phi_{\sigma=\pm 1}^{\rm (vac)}(t)=\pm\frac{1}{2}\left\{{\int_{0}^{t}\dot{\gamma}(t^{^{\prime
}})\left[1-\cos \lambda (t^{^{\prime }})\right]{\rm d}t^{^{\prime
}}}\right\}.                       \label{eq314}
\end{equation}

Investigation of quantum-vacuum geometric phases due to vacuum
fluctuation energies possesses theoretical significance: in
conventional {\it time-independent} quantum field theory the
infinite zero-point energy of vacuum is harmless and can be easily
removed by the normal-order procedure\cite{Bjorken2}. However, for
the {\it time-dependent} quantum field systems, ({\it e.g.},
photon fields propagating inside a helically curved fiber, and
quantum fields in an expanding universe or time-dependent
gravitational backgrounds), the time-dependent vacuum zero-point
fields may also participate in the time evolution process and
therefore cannot be regarded merely as an inactive onlooker ({\it
i.e.}, a simple passive background). According to the formulation
applied to the {\it time-independent} field theory, these
physically interesting vacuum effects would unfortunately have
been deducted by the second-quantization normal-order technique.
For this reason, we think that perhaps it is necessary to consider
the validity problem of normal-order procedure in {\it
time-dependent} quantum field theory.

Since quantum-vacuum geometric phases has an important connection
with vacuum quantum fluctuation, its experimental realization
deserves consideration, which will be briefly discussed at the end
of this section.
\subsection{Wave propagation in biaxially anisotropic left-handed
materials} It is well known that the geometric phases of photons
inside a curved fiber in previous experiments\cite{Tomita} often
depend on the cone angles of solid angles subtended by a curve
traced by the direction of wave vector of light at the center of
the photon momentum ${\bf k}$ space. Here, however, by taking into
account the peculiar properties of wave propagation in certain
biaxially anisotropic left-handed media, we will present a
physically interesting geometric phase that is independent of the
cone angles. First we consider the wave propagation in this type
of left-handed materials with the permittivity and permeability
tensors as follows
\begin{equation}
(\hat{\epsilon})_{ik}=\left(\begin{array}{cccc}
\epsilon  & 0 & 0 \\
0 &   -\epsilon & 0  \\
 0 &  0 &  \epsilon_{3}
 \end{array}
 \right),                 \qquad          (\hat{\mu})_{ik}=\left(\begin{array}{cccc}
-\mu  & 0 & 0 \\
0 &   \mu & 0  \\
 0 &  0 &  \mu_{3}
 \end{array}
 \right) .              \label{eq317}
\end{equation}
If the propagation vector of time-harmonic electromagnetic wave is
${\bf k}=(0, 0, k)$, then according to the Maxwellian Equations,
one can arrive at ${\bf k}\times {\bf E}=(-kE_{2}, kE_{1}, 0)$,
$[(\hat{\mu})_{ik}H_{k}]=(-\mu H_{1}, \mu H_{2}, 0)$. It follows
from the Faraday's electromagnetic induction law $\nabla
\times{\bf E}=-\frac{\partial{\bf B}}{\partial t}$ that
$H_{1}=\frac{kE_{2}}{\mu\mu_{0}\omega}$ and
$H_{2}=\frac{kE_{1}}{\mu\mu_{0}\omega}$. Thus, the third component
of Poynting vector of this time-harmonic wave is obtained
\begin{equation}
S_{3}=E_{1}H_{2}-E_{2}H_{1}=\frac{k}{\mu\mu_{0}\omega}(E_{1}^{2}-E_{2}^{2}),
\label{eq318}
\end{equation}
which implies that the Poynting vectors corresponding to the
$E_{1}$- and $E_{2}$- fields are of the form
\begin{equation}
{\bf S}^{(1)}=\frac{E_{1}^{2}}{\mu\mu_{0}\omega}{\bf k},    \quad
{\bf S}^{(2)}=-\frac{E_{2}^{2}}{\mu\mu_{0}\omega}{\bf k},
\label{eq319}
\end{equation}
respectively. It is apparently seen from (\ref{eq319}) that the
direction of wave vector ${\bf S}^{(2)}$ is opposite to that of
${\bf S}^{(1)}$. This, therefore, means that if $\mu>0$, then for
the $E_{1}$ field, this biaxially anisotropic medium characterized
by (\ref{eq317}) is like a right-handed material (regular
material) whereas for the $E_{2}$ field, it serves as a
left-handed one.

In view of above discussions, it is concluded that inside the
above biaxially anisotropic medium, the wave vectors of $E_{1}$-
and $E_{2}$- fields of propagating planar wave are apposite to
each other. Consider a hypothetical optical fiber that is made of
this biaxially anisotropic left-handed medium, inside which the
wave vector of $E_{1}$ field propagating is assumed to be ${\bf
k(t)}=k(\sin\lambda \cos\gamma, \sin\lambda \sin \gamma ,
\cos\lambda)$. If both $\lambda$ and $\gamma$ are nonvanishing,
then this fiber is noncoplanarly curved and in consequence the
geometric phases of light will arise. It is readily verified from
(\ref{eq319}) that the wave vector of $E_{2}$- field is $-{\bf
k(t)}=k(\sin\lambda' \cos\gamma', \sin\lambda' \sin \gamma' ,
\cos\lambda')$ with $\lambda'=\pi-\lambda$ and
$\gamma'=\gamma+\pi$. Note that for the latter case ({\it i.e.},
$E_{2}$- field), the expression for the time-dependent coefficient
in (\ref{eq224}) changes from
$\int_{0}^{t}\dot{\gamma}(t^{^{\prime }})\left[1-\cos \lambda
(t^{^{\prime }})\right]{\rm d}t^{^{\prime }}$ to
$\int_{0}^{t}\dot{\gamma}(t^{^{\prime }})\left[1+\cos \lambda
(t^{^{\prime }})\right]{\rm d}t^{^{\prime }}$ (because of
$\lambda\rightarrow\pi-\lambda$ and $\gamma\rightarrow\gamma+\pi$
). In the next subsection these results will be useful in
calculating the cone angle independent geometric phases of
circularly polarized light in biaxially anisotropic left-handed
media.
\subsection{Cone angle independent geometric phases in
biaxially anisotropic left-handed media}

The creation operators of left- and right- handed circularly
polarized light are
$a^{\dagger}_{L}=\frac{a^{\dagger}_{1}+ia^{\dagger}_{2}}{\sqrt{2}}$
and
$a^{\dagger}_{R}=\frac{a^{\dagger}_{1}-ia^{\dagger}_{2}}{\sqrt{2}}$,
respectively\cite{Bjorken}. The photon states corresponding to
right- and left- handed polarized light with the photon occupation
numbers being $n_{R}$ and $n_{L}$ are respectively defined to be
$|n_{R}\rangle=\frac{\left(a^{\dagger}_{R}\right)^{n_{R}}}{\sqrt{n_{R}!}}|0_{R}\rangle$
and
$|n_{L}\rangle=\frac{\left(a^{\dagger}_{L}\right)^{n_{L}}}{\sqrt{n_{L}!}}|0_{L}\rangle$.
Since according to the discussion in the previous subsection the
wave vector of $E_{1}$-field in such anisotropic left-handed media
is antiparallel to that of $E_{2}$-field, we should first
calculate the following expectation value $\langle n_{R}
|a_{1}^{\dagger}a_{1}|n_{R}\rangle$, $\langle n_{R}
|a_{2}^{\dagger}a_{2}|n_{R}\rangle$, $\langle n_{L}
|a_{1}^{\dagger}a_{1}|n_{L}\rangle$ and $\langle n_{L}
|a_{2}^{\dagger}a_{2}|n_{L}\rangle$ in order to obtain the
expressions for geometric phases of left- and right- handed
circularly polarized light in this peculiar biaxially anisotropic
left-handed medium. By the aid of
$a_{1}|n_{R}\rangle=\frac{1}{2^{\frac{n_{R}}{2}}\sqrt{n_{R}!}}\sum^{n_{R}}_{l=0}\frac{n_{R}!}{l!(n_{R}-l)!}a_{1}\left(a_{1}^{\dagger}\right)^{l}\left(ia_{2}^{\dagger}\right)^{n_{R}-l}|0_{R}\rangle$,
one can arrive at
\begin{eqnarray}
a_{1}|n_{R}\rangle&=&\frac{n_{R}}{2^{\frac{n_{R}}{2}}\sqrt{n_{R}!}}\sum^{n_{R}}_{l=1}\frac{\left(n_{R}-1\right)!}{(l-1)!(n_{R}-l)!}\left(a_{1}^{\dagger}\right)^{l-1}\left(ia_{2}^{\dagger}\right)^{n_{R}-l}|0_{R}\rangle
\nonumber \\
&=&\sqrt{\frac{n_{R}}{2}}\frac{1}{2^{\frac{n_{R}-1}{2}}\sqrt{\left(n_{R}-1\right)!}}\sum^{n_{R}}_{l=1}\frac{\left(n_{R}-1\right)!}{(l-1)!(n_{R}-l)!}\left(a_{1}^{\dagger}\right)^{l-1}\left(ia_{2}^{\dagger}\right)^{n_{R}-l}|0_{R}\rangle=\sqrt{\frac{n_{R}}{2}}|n_{R}-1\rangle,
\end{eqnarray}
where use is made of the formula
$a_{1}\left(a_{1}^{\dagger}\right)^{l}=l\left(a_{1}^{\dagger}\right)^{l-1}+\left(a_{1}^{\dagger}\right)^{l}a_{1}$.
Thus, we obtain
$a_{1}|n_{R}\rangle=\sqrt{\frac{n_{R}}{2}}|n_{R}-1\rangle$ and
consequently $\langle n_{R}
|a_{1}^{\dagger}=\sqrt{\frac{n_{R}}{2}}\langle n_{R}-1|$ and
$\langle n_{R}
|a_{1}^{\dagger}a_{1}|n_{R}\rangle=\frac{n_{R}}{2}$.

In the similar manner, one can obtain
\begin{equation}
a_{2}|n_{R}\rangle=i\sqrt{\frac{n_{R}}{2}}|n_{R}-1\rangle,  \quad
\langle n_{R} |a_{2}^{\dagger}=-i\sqrt{\frac{n_{R}}{2}}\langle
n_{R}-1| ,   \quad           \langle n_{R}
|a_{2}^{\dagger}a_{2}|n_{R}\rangle=\frac{n_{R}}{2}.
\end{equation}
Hence the nonadiabatic noncyclic geometric phases of right-handed
polarized photons corresponding to $E_{1}$- and $E_{2}$- fields
are
\begin{equation}
\phi_{R}^{(1)}(t)=\frac{n_{R}}{2}\left\{{\int_{0}^{t}\dot{\gamma}(t^{^{\prime
}})\left[1-\cos \lambda (t^{^{\prime }})\right]{\rm d}t^{^{\prime
}}}\right\},   \quad
\phi_{R}^{(2)}(t)=\frac{n_{R}}{2}\left\{{\int_{0}^{t}\dot{\gamma}(t^{^{\prime
}})\left[1+\cos \lambda (t^{^{\prime }})\right]{\rm d}t^{^{\prime
}}}\right\},
\end{equation}
respectively, and their sum is
\begin{equation}
\phi_{R}(t)=\phi_{R}^{(1)}(t)+\phi_{R}^{(2)}(t)=n_{R}\int_{0}^{t}\dot{\gamma}(t^{^{\prime
}}){\rm d}t^{^{\prime }},
\end{equation}
which is independent of the cone angle $\lambda (t)$ of photon
momentum $\bf k$ space.

In the same fashion, we obtain
\begin{equation}
a_{1}|n_{L}\rangle=\sqrt{\frac{n_{L}}{2}}|n_{L}-1\rangle,  \quad
\langle n_{L} |a_{1}^{\dagger}=\sqrt{\frac{n_{L}}{2}}\langle
n_{L}-1| ,   \quad           \langle n_{L}
|a_{1}^{\dagger}a_{1}|n_{L}\rangle=\frac{n_{L}}{2}
\end{equation}
and
\begin{equation}
a_{2}|n_{L}\rangle=-i\sqrt{\frac{n_{L}}{2}}|n_{L}-1\rangle,  \quad
\langle n_{L} |a_{2}^{\dagger}=i\sqrt{\frac{n_{L}}{2}}\langle
n_{L}-1| ,   \quad           \langle n_{L}
|a_{2}^{\dagger}a_{2}|n_{L}\rangle=\frac{n_{L}}{2}.
\end{equation}
Hence the nonadiabatic noncyclic geometric phases of left-handed
polarized photons corresponding to $E_{1}$- and $E_{2}$- fields
are
\begin{equation}
\phi_{L}^{(1)}(t)=-\frac{n_{L}}{2}\left\{{\int_{0}^{t}\dot{\gamma}(t^{^{\prime
}})\left[1-\cos \lambda (t^{^{\prime }})\right]{\rm d}t^{^{\prime
}}}\right\},         \quad
\phi_{L}^{(2)}(t)=-\frac{n_{L}}{2}\left\{{\int_{0}^{t}\dot{\gamma}(t^{^{\prime
}})\left[1+\cos \lambda (t^{^{\prime }})\right]{\rm d}t^{^{\prime
}}}\right\},
\end{equation}
respectively, and their sum is
\begin{equation}
\phi_{L}(t)=\phi_{L}^{(1)}(t)+\phi_{L}^{(2)}(t)=-n_{L}\int_{0}^{t}\dot{\gamma}(t^{^{\prime
}}){\rm d}t^{^{\prime }},
\end{equation}
which is also independent of the cone angle $\lambda (t)$.

Thus the total geometric phases of left- and right- handed
polarized photons is given by
\begin{equation}
\phi_{\rm tot}^{\rm (g)
}(t)=\phi_{R}(t)+\phi_{L}(t)=\left(n_{R}-n_{L}\right)\int_{0}^{t}\dot{\gamma}(t^{^{\prime
}}){\rm d}t^{^{\prime }},
\end{equation}
which differs from (\ref{eq312}) only by a cone angle $\lambda
(t)$ of photon momentum $\bf k$ space.

Since the geometric phases of both left- and right- handed
polarized light propagating in the above biaxially anisotropic
left-handed materials depend no longer on the cone angle, someone
may argue that the geometric phases presented here lose their
topological and global nature. This is not the true case.
Geometric phases presents the topological properties of quantum
systems in time-evolution process. Differing from dynamical phase
that depends on dynamical quantities of systems such as energy,
frequency, velocity as well as coupling coefficients, geometric
phase is independent of these dynamical quantities. Instead, it is
only related to the geometric nature of the pathway along which
quantum systems evolve. It follows that here $\phi_{\rm tot}^{\rm
(g)}(t)$ is related only to the precessional frequency
$\dot{\gamma}$ of photon propagation in the curved fiber, which is
not of dynamical nature and therefore cannot be considered the
dynamical quantity. This precessional frequency depends upon the
geometric shape of curved fiber. For example, in the helically
curved fiber, which was used first in Tomita and Chiao's
experiment to produce photon cyclic Berry's phase\cite{Tomita},
the precessional frequency equals $\frac{2\pi c}{\sqrt{d^{2}+(4\pi
a)^{2}}}$\cite{Shen1}, where $d$ and $a$ respectively denote the
pitch length and the radius of the helix, and $c$ is the speed of
light in a vacuum. For this reason, we think that $\phi_{\rm
tot}^{\rm (g)}(t)$ still possesses the geometric nature of time
evolution of photon wave function and can be regarded as the
geometric phase.
\subsection{Brief discussion: testing quantum-vacuum geometric phases}
As is stated above, the photon geometric phases at quantum-vacuum
level originates from the zero-point electromagnetic fluctuations.
Since geometric phases indicates topological and global properties
of quantum systems in time-evolution processes, the quantum-vacuum
geometric phases of electromagnetic fields in the helically wound
fiber may contain the information on the global properties of time
evolution of vacuum fluctuation fields. Moreover, we should make
it clear whether the normal-order procedure is valid or not in the
{\it time-dependent} quantum field theory and it is therefore
essential to detect the quantum-vacuum geometric phases
(\ref{eq314}) in experiments.

However, it should be pointed out that, unfortunately, even at the
quantum level, this observable quantum-vacuum geometric phases
$\phi_{\sigma=\pm 1}^{\rm (vac)}(t)$ is absent in the fiber
experiment, since it follows from (\ref{eq313}) and (\ref{eq314})
that the signs of quantal geometric phases of left- and
right-handed circularly polarized photons are just opposite to one
another, and so that the quantum-vacuum geometric phases would
have been counteracted by each other. Hence the observed geometric
phases are only those expressed by (\ref{eq312}), the adiabatic
case of which, as stated above, has been measured in the optical
fiber experiments performed by Tomita and Chiao {\it et
al.}\cite{Tomita,Kwiat,Ross,Haldane1}. It is impossible for
physicists to detect the quantum-vacuum geometric phases, which
has been eliminated, in these fiber experiments.

Since the quantum-vacuum geometric phases is so important but
unfortunately cancelled by each other (hence the total vacuum
geometric phases vanishes), we must ask such question: how can we
detect the quantum-vacuum geometric phases corresponding to {\it
only} one of the circularly polarized light? The studies in the
previous subsection enlighten us on this subject. If one can
design and fabricate a kind of such artificial composite
metamaterials, where for the left-handed polarized light the
material serves as a left-handed medium, while for the
right-handed polarized light it behaves like an ordinary material,
{\it i.e.}, the right-handed medium, then in this medium the
quantum-vacuum geometric phases corresponding to the right- and
left- handed polarized light are
\begin{equation}
  \phi_{R}^{\rm (vac)}(t)=\frac{1}{2}\left\{{\int_{0}^{t}\dot{\gamma}(t^{^{\prime
}})\left[1-\cos \lambda (t^{^{\prime }})\right]{\rm d}t^{^{\prime
}}}\right\},    \quad    \phi_{L}^{\rm
(vac)}(t)=-\frac{1}{2}\left\{{\int_{0}^{t}\dot{\gamma}(t^{^{\prime
}})\left[1+\cos \lambda (t^{^{\prime }})\right]{\rm d}t^{^{\prime
}}}\right\},                     \label{eq329}
\end{equation}
respectively. Here, the integrand in $\phi_{L}^{\rm (vac)}(t)$ is
obtained from $ \phi_{R}^{\rm (vac)}(t)$ via the parameter
replacements $\lambda\rightarrow \pi-\lambda$ and
$\gamma\rightarrow \gamma+\pi$.

Thus according to the expression (\ref{eq329}) that the total
quantum-vacuum geometric phases is $\phi_{\rm tot}^{\rm
(vac)}(t)=\phi_{L}^{\rm (vac)}(t)+\phi_{R}^{\rm
(vac)}(t)=-\int_{0}^{t}\dot{\gamma}(t^{^{\prime }})\cos \lambda
(t^{^{\prime }}){\rm d}t^{^{\prime }}$, which is no longer
vanishing in this artificial composite medium. Although it is of
physically interest to take into account this topic, designing
such materials is very complicated and is not the main subject in
this paper, so here we will not discuss further this problem. It
is under consideration and will be published elsewhere. Here we
only emphasize that it is truly possible for us to detect
quantum-vacuum geometric phases by using certain anisotropic
left-handed media. For instance, in section II, we stated that the
indices squared of left- and right- handed circularly polarized
light in gyrotropic media are respectively
$n_{L}^{2}=(\epsilon_{1}-\epsilon_{2})(\mu_{1}-\mu_{2})$ and
$n_{R}^{2}=(\epsilon_{1}+\epsilon_{2})(\mu_{1}+\mu_{2})$,
respectively\cite{Veselago}. If, for example, by taking some
certain values of $\epsilon_{1}$, $\epsilon_{2}$, $\mu_{1}$ and
$\mu_{2}$, then $n_{L}^{2}<0$ while $n_{R}^{2}>0$ and consequently
the left-handed polarized light cannot be propagated in this
medium, and in the meanwhile the quantum vacuum fluctuation
corresponding to the left-handed polarized light will also be
absorbed ({\it e.g.}, the wave amplitude exponentially decreases
because of the imaginary part of the refractive index) in this
anisotropic absorptive medium ({\it i.e.}, the vacuum-fluctuation
electromagnetic field alters its mode structures in the absorptive
medium). For this reason, the only retained geometric phases is
that of right-handed polarized light, which we can test
experimentally.
\section{Extra phases of light at the interfaces between left- and right- handed media}
In this section, we consider the effects of light appearing at the
interfaces between left- and right- handed media. In order to
treat this problem conveniently, we study the wave propagation
inside an optical fiber which is periodically modulated by
altering regular and negative media. Although it is doubtful
whether such periodically modulated fibers could be designed and
realized or not in experiments at the optical scale, it could be
argued that the work presented here can be considered only a
speculative one. But the method and results obtained via the use
of this optical fiber system composed by such sequences of right-
and left- handed materials can also be applied to the light
propagation at the interfaces between left- and right- handed
media in arbitrary geometric shapes of optical materials. In this
periodical optical fiber, helicity inversion (or the transitions
between helicity states) of photons may be easily caused by the
interaction of light field with media near both sides of the
interfaces between LRH materials. Since photon helicity inversion
at the interfaces mentioned above is a {\it time-dependent}
process, this new geometric phase arises during the light
propagates through the interfaces (in the following we will call
them the LRH interfaces) between left- and right- handed media,
where the anomalous refraction occurs when the incident lightwave
travels to the LRH interfaces. we think that, in the literature it
gets less attention than it deserves. In what follows we calculate
the photon wavefunction and corresponding extra phases (including
the geometric phases) in this physical process, and emphasize that
we should attach importance to this geometric phases when
considering the wave propagation near the LRH interfaces.
\subsection{Model Hamiltonian}
We now treat the helicity reversal problem of light wave adjacent
to the interfaces of left- and right- handed media. For
convenience, let us consider a hypothetical optical fiber that is
fabricated periodically from both left- and right- handed (LRH)
media with the optical refractive indices being $-n$ and $n$,
respectively. Thus the wave vector of photon moving along the
fiber is respectively $-n\frac{\omega }{c}$ in left-handed (LH)
section and $n\frac{\omega }{c}$ in right-handed (RH) section,
where $\omega$ and $c$ respectively denote the frequency and the
speed of light in a vacuum. For simplicity, we assume that the
periodical length, $b$, of LH is equal to that of RH in the fiber.
If the eigenvalue of photon helicity is $\sigma $ in right-handed
sections, then, according to the definition of helicity,
$h=\frac{\bf k}{\left| {\bf k}\right| }\cdot{\bf J}$ with ${\bf
J}$ denoting the total angular momentum of the photon, the
eigenvalue of helicity acquires a minus sign in left-handed
sections. We assume that at $t=0$ the light propagates in the
right-handed section and the initial eigenvalue of photon helicity
is $\sigma $. So, in the wave propagation inside the
LRH-periodical optical fiber, the helicity eigenvalue of $h$ is
then $\left(-\right) ^{m}\sigma $ with $m=\left[
\frac{ct}{nb}\right] $, where $\left[\frac{ct}{nb}\right] $
represents the integer part of $\frac{ct}{nb}$. It is clearly seen
that $\left(-\right) ^{m}$ stands for the switching on and off of
the helicity reversal, {\it i.e.}, the positive and negative value
of $\left(-\right) ^{m}$ alternate in different time intervals.
This, therefore, means that if $2k\left(\frac{nb}{c}\right) <t\leq
\left(2k+1\right) \left( \frac{nb}{c}\right) $, then
$\left(-\right) ^{m} =+1$, and if $\left( 2k+1\right)
\left(\frac{nb}{c}\right) <t\leq (2k+2)\left( \frac{nb}{c}\right)
$, then $\left(-\right) ^{m} =-1$, where $k$ is zero or a positive
integer. It follows that the incidence of lightwave on the LRH
interfaces in the fiber gives rise to the transitions between the
photon helicity states ($\left| +\right\rangle $ and $\left|
-\right\rangle $). This enables us to construct a time-dependent
effective Hamiltonian
\begin{equation}
H\left(t\right) =\frac{1}{2}\omega \left(t\right) \left(
S_{+}+S_{-}\right)                     \label{eq1}
\end{equation}
in terms of $\left| +\right\rangle $ and $\left| -\right\rangle $
to describes this instantaneous transition process of helicity
states at the LRH interfaces, where $S_{+}=\left| +\right\rangle
\left\langle -\right| ,S_{-}=\left| -\right\rangle \left\langle
+\right| ,S_{3}=\frac{1}{2}\left(\left| +\right\rangle
\left\langle +\right| -\left| -\right\rangle \left\langle -\right|
\right) $ satisfying the following SU(2) Lie algebraic commuting
relations $\left[ S_{+},S_{-}\right] =2S_{3}$ and $\left[
S_{3},S_{\pm }\right] =\pm S_{\pm }$. The time-dependent frequency
parameter $\omega \left(t\right) $ may be taken to be $\omega
\left( t\right) =\varsigma \frac{\rm d}{{\rm d}t}p\left(t\right)
$, where $p\left(t\right) =\left(-\right) ^{m}$ with $m=\left[
\frac{ct}{nb}\right] $, and $\varsigma$ is the coupling
coefficient, which can, in principle, be determined by the
physical mechanism of interaction between light fields and media.
Since $p\left(t\right) $ is a periodical function, by using the
analytical continuation procedure, it can be rewritten as the
following linear combinations of analytical functions
\begin{equation}
p\left(t\right) =\sum_{k=1}^{\infty }\frac{2}{k\pi }\left[
1-\left(-\right) ^{k}\right] \sin \left(\frac{k\pi c}{nb}t\right).
\label{eq2}
\end{equation}

 In what follows, we solve the time-dependent Schr\"{o}dinger
equation (in the unit $\hbar =1$)
\begin{equation}
H\left(t\right) \left| \Psi _{\sigma }\left(t\right) \right\rangle
=i\frac{\partial }{\partial t}\left| \Psi _{\sigma }\left(
t\right) \right\rangle     \label{eq3}
\end{equation}
governing the propagation of light in the LRH- periodical fiber.
According to the Lewis-Riesenfeld invariant theory\cite{Lewis},
the exact particular solution $\left| \Psi _{\sigma
}\left(t\right) \right\rangle$ of the time-dependent
Schr\"{o}dinger equation (\ref{eq3}) is different from the
eigenstate of the invariant $I(t)$ only by a time-dependent $c$-
number factor $\exp \left[ \frac{1}{i}\phi _{\sigma
}\left(t\right) \right]$, where
\begin{equation}
\phi _{\sigma }\left(t\right)=\int_{0}^{t}\left\langle \Phi
_{\sigma }\left(t^{\prime }\right) \right|[H(t^{\prime
})-i\frac{\partial }{\partial t^{\prime }}]\left| \Phi _{\sigma
}\left(t^{\prime }\right) \right\rangle {\rm d}t^{\prime }
 \label{eq4}
\end{equation}
with $\left| \Phi _{\sigma }\left(t\right) \right\rangle $ being
the eigenstate of the invariant $I(t)$ (corresponding to the
particular eigenvalue $\sigma$) and satisfying the eigenvalue
equation $I\left(t\right) \left| \Phi _{\sigma }\left(t\right)
\right\rangle =\sigma \left| \Phi _{\sigma }\left(t\right)
\right\rangle$, where the eigenvalue $\sigma$ of the invariant
$I(t)$ is {\it time-independent}. Thus we have
\begin{equation}
\left| \Psi _{\sigma }\left(t\right) \right\rangle =\exp \left[
\frac{1}{i}\phi _{\sigma }\left(t\right) \right] \left| \Phi
_{\sigma }\left(t\right) \right\rangle.    \label{eq51}
\end{equation}
In order to obtain $\left| \Psi _{\sigma }\left(t\right)
\right\rangle$, we should first obtain the eigenstate $\left| \Phi
_{\sigma }\left(t\right) \right\rangle $ of the invariant $I(t)$.
\subsection{Photon geometric phases due to helicity inversions inside a periodical fiber made of Left-handed media}
Here we investigate the time evolution of photon wavefunctions and
extra phases due to photon helicity inversion at the LRH
interfaces. As has been stated above, for convenience, we consider
the wave propagation inside a hypothetical optical fiber which is
composed periodically of left- and right- handed media. Now we
solve the time-dependent Schr\"{o}dinger equation. In accordance
with the Lewis-Riesenfeld theory, the invariant $I(t)$ is a
conserved operator ({\it i.e.}, it possesses {\it
time-independent} eigenvalues) and agrees with the following
Liouville-Von Neumann equation

\begin{equation}
\frac{\partial I(t)}{\partial t}+\frac{1}{i}[I(t),H(t)]=0.
                 \label{eq5}
\end{equation}
It follows from Eq.(\ref{eq5}) that the invariant $I(t)$ may also
be constructed in terms of $S_{\pm }$ and $S_{3}$, {\it i.e.},

\begin{equation}
I\left(t\right) =2\left\{\frac{1}{2}\sin \theta \left(t\right)
\exp \left[ -i\varphi  \left(t\right)\right] S_{+}+\frac{1}{2}\sin
\theta \left(t\right) \exp \left[ i\varphi \left(t\right)\right]
S_{-}+\cos \theta \left(t\right) S_{3}\right\}. \label{eq6}
\end{equation}
Inserting Eq.(\ref{eq1}) and (\ref{eq6}) into Eq.(\ref{eq5}), one
can arrive at a set of auxiliary equations

\begin{eqnarray}
\exp \left[ -i\varphi \right] \left(\dot{\theta}\cos \theta
-i\dot{\varphi}\sin \theta \right) -i\omega \cos \theta =0,  \nonumber \\
\quad \dot{\theta}+\omega \sin \varphi =0,
 \label{eq7}
\end{eqnarray}
which are used to determine the time-dependent parameters,
$\theta\left(t\right)$ and $\varphi\left(t\right)$, of the
invariant $I(t)$\cite{Lewis}.

It should be noted that we cannot easily solve the eigenvalue
equation $I\left(t\right) \left| \Phi _{\sigma }\left(t\right)
\right\rangle =\sigma \left| \Phi _{\sigma }\left(t\right)
\right\rangle$, for the time-dependent parameters $\theta\left(
t\right)$ and $\varphi\left(t\right)$ are involved in the
invariant (\ref{eq6}). If, however, we could find (or construct) a
unitary transformation operator $V(t)$ to make $V^{\dagger
}(t)I(t)V(t)$ be {\it time-independent}, then the eigenvalue
equation problem of $I(t)$ is therefore easily resolved. According
to our experience for utilizing the invariant-related unitary
transformation formulation\cite{Shen2}, we suggest a following
unitary transformation operator

\begin{equation}
V\left(t\right) =\exp \left[ \beta \left(t\right) S_{+}-\beta
^{\ast }\left(t\right) S_{-}\right] ,
 \label{eq8}
\end{equation}
where $\beta (t)$ and $\beta ^{\ast }(t)$ will be determined by
calculating $I_{\rm V}=V^{\dagger }(t)I(t)V(t)$ in what follows.

Calculation of $I_{\rm V}=V^{\dagger }(t)I(t)V(t)$ yields

\begin{equation}
I_{\rm V}=V^{\dagger }\left(t\right) I\left(t\right) V\left(
t\right) =2S_{3},
 \label{eq9}
\end{equation}
if $\beta $ and $\beta ^{\ast }$ are chosen to be $\beta \left(
t\right) =-\frac{\theta \left(t\right) }{2}\exp \left[ -i\varphi
\left(t\right) \right]$, $\beta ^{\ast }\left(t\right)
=-\frac{\theta \left(t\right) }{2}\exp \left[ i\varphi \left(
t\right) \right] $. This, therefore, means that we can change the
{\it time-dependent} $I(t)$ into a {\it time-independent} $I_{\rm
V}$, and the result is $I_{\rm V}=2S_{3}$. Thus, the eigenvalue
equation of $I_{\rm V}$ is $I_{\rm V}\left| \sigma \right\rangle
=\sigma \left| \sigma \right\rangle $ with $\sigma=\pm 1$, and
consequently the eigenvalue equation of $I(t)$ is written $I\left(
t\right) V\left(t\right) \left| \sigma \right\rangle =\sigma
V\left(t\right) \left| \sigma \right\rangle $. So, we obtain the
eigenstate $\left| \Phi _{\sigma }\left(t\right) \right\rangle$ of
$I(t)$, {\it i.e.}, $\left| \Phi _{\sigma }\left(t\right)
\right\rangle=V\left(t\right)\left| \sigma \right\rangle$.

Correspondingly, $H(t)$ is transformed into

\begin{eqnarray}
H_{\rm V}\left(t\right) =V^{\dagger }\left(t\right) \left[ H\left(
t\right) -i\frac{\partial }{\partial t}\right] V\left(t\right)
\label{eq11}
\end{eqnarray}
and the time-dependent Schr\"{o}dinger equation (\ref{eq3}) is
rewritten
\begin{equation}
H_{\rm V}\left(t\right) \left| \Psi _{\sigma }\left(t\right)
\right\rangle_{\rm V} =i\frac{\partial }{\partial t}\left| \Psi
_{\sigma }\left(t\right) \right\rangle_{\rm V}   \label{eq12}
\end{equation}
under the unitary transformation $V(t)$, where $ \left| \Psi
_{\sigma }\left(t\right) \right\rangle_{\rm V}=V^{\dagger }\left(
t\right) \left| \Psi _{\sigma }\left(t\right) \right\rangle$.

Further analysis shows that the exact particular solution $ \left|
\Psi _{\sigma }\left(t\right) \right\rangle_{\rm V}$ of the
time-dependent Schr\"{o}dinger equation (\ref{eq12}) is different
from the eigenstate $\left| \sigma \right\rangle$ of the {\it
time-independent} invariant $I_{\rm V}$ only by a time-dependent
$c$- number factor $\exp \left[ \frac{1}{i}\phi _{\sigma }\left(
t\right) \right]$\cite{Lewis}, which is now rewritten as $\exp
\left \{ \int_{0}^{t}\left\langle \sigma\right| H_{\rm V}\left(
t^{\prime }\right)\left|\sigma \right\rangle {\rm d}t^{\prime
}\right\}$.

By using the auxiliary equations (\ref{eq7}), the Glauber formula
and the Baker-Campbell-Hausdorff formula\cite{Wei}, it is verified
that $H_{\rm V}(t)$ depends only on the operator $S_{3}$, {\it
i.e.},

\begin{equation}
H_{\rm V}\left(t\right) =\left\{ \omega \left(t\right)\sin \theta
\left(t\right) \cos \varphi \left(t\right)+\dot{\varphi}\left(
t\right) \left[ 1-\cos \theta \left(t\right) \right] \right\}
S_{3}
 \label{eq13}
\end{equation}
and the time-dependent $c$- number factor $\exp \left[
\frac{1}{i}\phi _{\sigma }\left(t\right) \right] $ is therefore
$\exp \left\{ \frac{1}{i}\left[ \phi _{\sigma }^{\left({\rm
d}\right) }\left(t\right) +\phi _{\sigma }^{\left({\rm g}\right)
}\left(t\right) \right] \right\} $, where the dynamical phase is

\begin{equation}
\phi _{\sigma }^{\left({\rm d}\right) }\left(t\right) =\sigma
\int_{0}^{t}\omega \left(t^{\prime }\right)\sin \theta \left(
t^{\prime }\right) \cos \varphi \left(t^{\prime }\right){\rm
d}t^{\prime } \label{eq14}
\end{equation}
and the geometric phase is
\begin{equation}
\phi _{\sigma }^{\left({\rm g}\right) }\left(t\right) =\sigma
\int_{0}^{t}\dot{\varphi}\left(t^{\prime }\right) \left[ 1-\cos
\theta \left(t^{\prime }\right) \right] {\rm d}t^{\prime }.
 \label{eq15}
\end{equation}

Hence the particular exact solution of the time-dependent
Schr\"{o}dinger equation (\ref{eq3}) corresponding to the
particular eigenvalue, $\sigma$, of the invariant $I(t)$ is of the
form

\begin{equation}
\left| \Psi _{\sigma }\left(t\right) \right\rangle =\exp \left\{
\frac{1}{i}\left[ \phi _{\sigma }^{\left({\rm d}\right) }\left(
t\right) +\phi _{\sigma }^{\left({\rm g}\right) }\left(t\right)
\right] \right\} V\left(t\right) \left| \sigma \right\rangle.
 \label{eq16}
\end{equation}

It follows from the obtained expression (\ref{eq15}) for geometric
phase of photons that, if the frequency parameter $\omega$ is
small ({\it i.e.}, the adiabatic quantum process) and then
according to the auxiliary equations (\ref{eq7}),
$\dot{\theta}\simeq 0$, the Berry phase (adiabatic geometric phase
) in a cycle ({\it i.e.}, one round trip, $T\simeq
\frac{2\pi}{\omega}$) of parameter space of invariant $I(t)$ is

\begin{equation}
\phi _{\sigma }^{\left({\rm g}\right) }\left(T\right)= 2\pi \sigma
(1-\cos \theta),
 \label{eq17}
\end{equation}
where $2\pi (1-\cos \theta)$ is a solid angle over the parameter
space of the invariant $I(t)$, which means that the geometric
phase is related only to the geometric nature of the pathway along
which quantum systems evolve. Expression (\ref{eq17}) is analogous
to the magnetic flux produced by a monopole of strength $\sigma$
existing at the origin of the parameter space. This, therefore,
implies that geometric phases differ from dynamical phases and
involve the global and topological properties of the time
evolution of quantum systems.
\subsection{In biaxially anisotropic left-handed media}
Note that in the previous subsection, we treat the helicity
reversals of single photon in electromagnetic media made of
isotropic left- and right- handed materials. Now we consider this
problem in a system composed by a sequence of right-handed
(isotropic) and biaxially anisotropic left-handed media, the
permittivity and permeability of the latter is given in
(\ref{eq317}). It has been verified in Sec.III that for the
$E_{1}$- field this biaxially anisotropic medium can be regarded
as a right-handed material while for the $E_{2}$- field it can be
considered a left-handed one. This, therefore, implies that only
the $E_{2}$- field propagating through this biaxially anisotropic
left-handed medium will acquire an additional phase due to its
helicity inversion at the LRH interfaces. But since the $E_{2}$-
field is not the eigenmode of the photon helicity, we cannot
obtain the extra phases immediately by using the formula
(\ref{eq4}). According to the treatment in Sec.III, one can arrive
at the expression for the total extra phases of circularly
polarized light (with the occupation numbers of polarized photons
being $n_{L}$ and $n_{R}$) propagating through the system made of
a sequence of isotropic regular media and biaxially anisotropic
left-handed one, and the result is written as
\begin{equation}
\phi_{\rm tot}(t)=\frac{n_{R}-n_{L}}{2}\left[\phi^{\rm
(d)}(t)+\phi^{\rm (g)}(t)\right]
\end{equation}
with $\phi^{\rm (d)}(t)+\phi^{\rm
(g)}(t)=\int_{0}^{t}\left\{\omega \left(t^{\prime }\right)\sin
\theta \left(t^{\prime }\right) \cos \varphi \left(t^{\prime
}\right)+\dot{\varphi}\left(t^{\prime }\right) \left[ 1-\cos
\theta \left(t^{\prime }\right) \right]\right\}{\rm d}t^{\prime
}$. It follows that if $n_{R}=n_{L}$, the added phase due to
photon helicity reversal on the interfaces between left- and
right- handed media vanishes.

Based on the restriction Eq.(\ref{eqq27}) imposed on the frequency
shifts of circularly polarized light, we now consider a
possibility that only one of the polarized light, say $E_{R}$, can
be propagated in biaxially gyrotropic left-handed media. If the
frequency shift $\Omega_{L}$ of left-handed polarized light is
$-\omega-i\Gamma$, {\it i.e.}, $\omega+\Omega_{L}=-i\Gamma$, where
$\Gamma$ is a positive real number, then according to
Eq.(\ref{eqq27}), one can arrive at
\begin{equation}
n^{4}(\Gamma^{2}+\omega^{2})(2\omega+\Omega_{R})\Omega_{R}+\zeta^{4}\Gamma^{2}(\omega+\Omega_{R})^{2}=0.
\label{eq419}
\end{equation}
It is easy to obtain $\Omega_{R}$ from Eq.(\ref{eq419}), and the
result is given
$\Omega_{R}=-\omega+\gamma^{\frac{1}{2}}(1+\gamma)^{-\frac{1}{2}}\omega$
with
$\gamma=\frac{n^{4}(\Gamma^{2}+\omega^{2})}{\zeta^{4}\Gamma^{2}}$.
Thus, the frequency of right-handed polarized light is
\begin{equation}
\omega+\Omega_{R}=\gamma^{\frac{1}{2}}(1+\gamma)^{-\frac{1}{2}}\omega,
\label{eq420}
\end{equation}
which means that the frequency of $E_{R}$ is modified by a factor
$\gamma^{\frac{1}{2}}(1+\gamma)^{-\frac{1}{2}}$. If no L-R
coupling exists, {\it i.e.}, $\zeta=0$, then $\gamma$ tends to
infinity and the factor
$\gamma^{\frac{1}{2}}(1+\gamma)^{-\frac{1}{2}}$ approaches unity,
and then the frequency shift $\Omega_{R}$ of right-handed
polarized light is vanishing, which can be easily seen in the
expression (\ref{eq420}).

Note that in the case discussed above, the left-handed polarized
light in this type of media exponentially decreases (due to the
imaginary frequency $\omega+\Omega_{L}$, which is $-i\Gamma$)
while the right-handed one can be propagated, {\it i.e.}, only one
wave can be present in this media. So, in this case the additional
phase acquired by photon wavefunction due to helicity inversions
on LRH interfaces is
\begin{equation}
\phi_{\rm tot}(t)=\frac{n_{R}}{2}\left[\phi^{\rm (d)}(t)+\phi^{\rm
(g)}(t)\right].
\end{equation}

Additionally, it is of interest to show that the above scheme is
applicable to the detection of quantum-vacuum geometric phases
expressed by (\ref{eq313}). Since the left-handed polarized light
(including the quantum vacuum fluctuation corresponding to the
left-handed polarized light) cannot be propagated in this
media\cite{Klimov,Veselago}, the quantum-vacuum geometric phase of
right-handed polarized light will not be cancelled by that of
left-handed one, namely, the only retained quantum-vacuum
geometric phase is that of right-handed circularly polarized light
and therefore it is possible for the nonvanishing quantum-vacuum
geometric phases to be detected in experiments.
\subsection{Discussion: physical significance and potential applications}
It is worthwhile to point out that the geometric phase of photons
due to helicity inversion presented here is of quantum level.
However, whether the Chiao-Wu geometric phase due to the spatial
geometric shape of fiber is of quantum level or not is not
apparent (see, for example, the arguments between Haldane and
Chiao {\it et al.} about this problem \cite{Haldane2,Chiao2}),
since the expression for the Chiao-Wu geometric phase can be
derived by using both the classical Maxwell's electromagnetic
theory, differential geometry and quantum
mechanics\cite{Kwiat,Robinson,Haldane1,Haldane2,Chiao2}. However,
the geometric phase in this paper can be considered only by
Berry's adiabatic quantum theory and Lewis-Riesenfeld invariant
theory, namely, the classical electrodynamics cannot predict this
geometric phase. Although many investigators have taken into
account the boundary condition problem and anomalous refraction in
left-handed media by using the classical Maxwell's
theory\cite{Shelby,Veselago}, less attention is paid to this
geometric phase due to helicity inversion. It is believed that
this geometric phase originates at the quantum level, but survives
the correspondence-principle limit into the classical level. So,
We emphasize that it may be essential to take into consideration
this geometric phase in investigating the anomalous refraction at
the LRH interfaces.

It is well known that geometric phases arise only in
time-dependent quantum systems. In the present problem, the
transitions between helicity states on the LRH interfaces, which
is a time-dependent process, results in the geometric phase of
photons. This may be viewed from two aspects: (i) it is apparently
seen in Eq.(\ref{eq7}) that if the frequency parameter $\omega$ in
the Hamiltonian (\ref{eq1}) vanishes, then $\dot{\varphi}=0$ and
the geometric phase (\ref{eq15}) is therefore vanishing; (ii) it
follows from (\ref{eq2}) that the frequency coefficient $\omega
\left(t\right)$ of Hamiltonian (\ref{eq1}) is

\begin{equation}
\omega \left(t\right) =\varsigma \frac{\rm d}{{\rm d}t}p\left(
t\right) =\frac{2c}{nb}\varsigma\sum_{k=1}^{\infty }\left[
1-\left(-\right) ^{k}\right] \cos \left(\frac{k\pi c}{nb}t\right).
 \label{eq18}
\end{equation}
Since ${\left| \cos \left(\frac{k\pi c}{nb}t\right)\right|\leq
1}$, the frequency coefficient, the transition rates between
helicity states, and the consequent time-dependent phase ($\varphi
_{\sigma }^{\left({\rm g}\right) }\left(t\right)+\varphi _{\sigma
}^{\left({\rm d}\right) }\left(t\right)$) greatly decrease
correspondingly as the periodical optical path $nb$ increases.
Thus we can conclude that the interaction of light fields with
media near the LRH interfaces gives rise to this topological
quantum phase.

In addition to obtaining the expression (\ref{eq15}) for geometric
phase, we obtain the wavefunction (\ref{eq16}) of photons in the
LRH- optical fiber by solving the time-dependent Schr\"{o}dinger
equation (\ref{eq3}) based on the Lewis-Riesenfeld invariant
theory\cite{Lewis} and the invariant-related unitary
transformation formulation\cite{Gao,Gao2}. We believe that this
would enable us to consider the propagation of light fields inside
the optical fiber in more detail.

In the above treatment, we constructed an effective Hamiltonian
(\ref{eq1}) to describe the time evolution of helicity states of
photons. It should be noted that the method presented here is only
a phenomenological description of propagation of lightwave in the
LRH-periodical fiber. This phenomenological description is based
on the assumption that the direction of wave vector ${\bf k}$
becomes opposite nearly instantaneously on the LRH interfaces.
This assumption holds true so long as the periodical length $b$ is
much larger than the wavelength of lightwave in the fiber.
\\ \\

To close this section, we conclude with some remarks on the
potential significance of the subject in this section:

(i) The obtained geometric phase itself is physically interesting.
Moreover, it is necessary to consider this geometric phase in
discussing the anomalous refraction and wave propagation in
left-handed media (adjacent to the LRH interfaces).

(ii) Helicity inversion of photons, which is in exact analogy with
the transition operation between $0$ and $1$ in digital circuit,
can be caused due to the electromagnetic interactions at LRH
interfaces, and the time evolution of helicity states is governed
by (\ref{eq16}). It is of essential significance to control and
utilize the degrees of freedom of photons (photon number,
polarization, helicity, geometric phase, {\it etc}.) in
information science and technology\cite{Guo}. In the curved
optical fiber, the interaction of the photon spin with the wave
vector causes the helicity inversion of the photon, some authors
have considered its application to information
theory\cite{Guo,Muller}. Likewise, here we think the instantaneous
process of photon helicity reversals may also have some potential
applications in information technology and therefore deserve
further investigation.

(iii) It is of physical interest to consider the quantum effects
such as propagation of photons field, polarization of photon
states (time evolution of photon wavefunction) and spontaneous
emission decay rate of atoms\cite{Klimov} in left-handed media. In
this section, an illustrative example of quantum effects of
photons field resulting from helicity inversion caused by the LRH
interfaces is presented. We hope the present consideration in this
paper would open up new opportunities for investigating more
quantum mechanical properties, phenomena and effects in
left-handed media.
\section{Concluding remarks}
We consider the bianisotropic structures in optical ``constants''
of left-handed media whereby the left-right coupling of circularly
polarized light and cone angle independent geometric phases
arises.

Nonadiabatic noncyclic geometric phases of photons in a
noncoplanarly curved optical fiber fabricated from biaxially
anisotropic left-handed media is considered in this paper. It is
well known that the geometric phases of photons inside a curved
fiber in previous experiments often relate to the cone angles of
solid angles subtended at the center. Here, however, we present a
new geometric phase that is independent of the cone angles, by
taking into account the peculiar properties of wave propagation in
some anisotropic left-handed media. Some related topics ({\it
i.e.}, experimental realizations of {\it quantum vacuum geometric
phases} of photons by using certain anisotropic ({\it e.g.},
gyrotropic) left-handed materials, and {\it photon geometric
phases due to instantaneous helicity inversion} on the interfaces
between left- and right- handed media) are also discussed.

Geometric phases of circularly polarized light has some possible
applications to quantum computation\cite{Wangzd,Wangxb}, since the
wave propagation of coiled light is somewhat analogous to the
behavior of nuclear magnetic resonance system\cite{Wangxb}.
Realizing quantum computation by means of geometric origin of
adiabatic cyclic Berry's phase is now receiving attention due to
its intrinsic tolerance to noise. Recently, a conditional
geometric phase shift gate, which is fault tolerant to certain
types of errors due to its geometric nature, was realized via
nuclear magnetic resonance (NMR) under {\it adiabatic}
conditions\cite{Jones}. However, the {\it adiabatic} conditions
makes any fast conditional Berry's phase shift impossible. So,
more recently, Wang and Keiji suggested a scheme of {\it
nonadiabatic} conditional geometric phase shift with
NMR\cite{Wangxb}, the mathematical treatment of which is just the
same as (\ref{eq221})-(\ref{eq224}), so long as both the cone
angle $\lambda$ and the precessional frequency $\dot{\gamma}$
(denoted by $\Omega$) are taken to be constant.
 \\

To summarize, in this paper we focus on some interesting
properties of wave propagation of polarized photons in biaxially
anisotropic left-handed materials, which has not been considered
yet by the authors in the fields of classical optics, condensed
matter physics and materials science. Since it is possible for
current technology to fabricate uniaxially and biaxially
anisotropic left-handed media in experiments, we hope the optical
effects and phenomena discussed above would be investigated
experimentally in the near future.
\\ \\

\textbf{Acknowledgements}  I thank Sai-Ling He and Xiao-Chun Gao
for their helpful suggestions respectively on the wave propagation
in left-handed media and geometric phases in optical fibers. This
project is supported by the National Natural Science Foundation of
China under the project No. $90101024$.
\\ \\

\textbf{Appendices}
\\

{\bf Appendix I}. Three derivations of effective Hamiltonian of
photons inside a noncoplanarly curved fiber
\\

The effective Hamiltonian describing the light wave propagation in
a curved optical fiber is helpful in considering the nonadiabatic
noncyclic time evolution process of photon wavefunction in the
fiber. We have three methods to derive this effective Hamiltonian
(\ref{eq220}).
\\

{\bf Method i} \quad   By using the infinitesimal rotation
operator of wavefunction

The photon wavefunction $|\sigma, {\bf k(t)}\rangle$ varies as it
rotates by an infinitesimal angle, say $\vec{\vartheta}$, namely,
it obeys the following transformation rule
\begin{equation}
|\sigma, {\bf k'(t)}\rangle=\exp\left[-i\vec{\vartheta}\cdot{\bf
J}\right]|\sigma, {\bf k(t)}\rangle,       \eqnum{A1} \label{eqA1}
\end{equation}
where $\exp\left[-i\vec{\vartheta}\cdot{\bf J}\right]\simeq
1-i\vec{\vartheta}\cdot{\bf J}$ with ${\bf J}$ being the total
angular momentum operator of photon and ${\bf k'(t)}={\bf
k(t)}+\Delta {\bf k(t)}$ with $\Delta {\bf
k(t)}=\dot{\bf{k}}{\Delta t}$. Here $|\vec{\vartheta}|$ is the
angle between ${\bf k(t)}$ and ${\bf k'(t)}$, and the direction of
$\vec{\vartheta}$ is parallel to that of ${\bf k(t)}\times{\bf
k'(t)}$. One can therefore arrive at
\begin{equation}
\vec{\vartheta}=\frac{{\bf k(t)}\times{\bf k'(t)}}{k^2}=\frac{{\bf
k(t)}\times\dot{\bf{k}}}{k^2}{\Delta t}.
 \eqnum{A2} \label{eqApp2}
\end{equation}
Thus it follows from Eq.(\ref{eqA1}) and (\ref{eqApp2}) that
\begin{equation}
i\frac{\partial \left| \sigma ,{\bf{k}}(t)\right\rangle }{\partial
t}=\frac{{\bf{k}}(t)\times \dot{\bf{k}}(t)}{k^{2}}\cdot
{\bf{J}}\left| \sigma ,{\bf{k}}(t)\right\rangle      \eqnum{A3}
\label{eqA3}
\end{equation}
by calculating the time derivative of $|\sigma, {\bf
k'(t)}\rangle$. The total angular momentum is
${\bf{J}}={\bf{L}}+{\bf{S}}$, where the orbital angular momentum
${\bf{L}}$ is orthogonal to the linear momentum ${\bf{k}}$ for the
photon. So, $\frac{{\bf{k}}(t)\times \dot{\bf{k}}(t)}{k^{2}}\cdot
{\bf{L}}=0$ and the only retained term in $\frac{{\bf{k}}(t)\times
\dot{\bf{k}}(t)}{k^{2}}\cdot {\bf{J}}$ is $\frac{{\bf{k}}(t)\times
\dot{\bf{k}}(t)}{k^{2}}\cdot {\bf{S}}$. This, therefore, means
that if we think of Eq.(\ref{eqA3}) as the time-dependent
Schr\"{o}dinger equation governing the propagation of photons in
the noncoplanar fiber, then we can obtain the effective
Hamiltonian (\ref{eq220}).
\\

{\bf Method ii} \quad  By using the equation of motion of a photon

If the momentum squared ${\bf k}^2$ of a photon moving in a
noncoplanarly curved optical fiber is conserved, then we can
derive the following identity
\begin{equation}
\dot{\bf{k}}+{\bf{k}}\times (\frac{{\bf{k}}\times
\dot{\bf{k}}}{k^{2}})=0,             \eqnum{A4} \label{A4}
\end{equation}
which can be regarded as the equation of motion of a photon in the
fiber. Since Eq.(\ref{A4}) is exactly analogous to the equations
of motion of a charged particle moving in a magnetic field or a
spinning particle moving in a rotating frame of reference,
$-\frac{{\bf{k}}\times \dot{\bf{k}}}{k^{2}}$ can be considered a
``magnetic field'' or ``gravitomagnetic field'' (thus
$-{\bf{k}}\times (\frac{{\bf{k}}\times \dot{\bf{k}}}{k^{2}})$ can
be thought of as a ``Lorentz magnetic force'' or ``Coriolis
force''). Similar to the Mashhoon {\it et al.}'s work ({\it i.e.},
the derivation of the interaction Hamiltonian of gravitomagnetic
dipole moment in a gravitomagnetic field)\cite{Shen2,Mashhoon},
one can also readily write the Hamiltonian describing the coupling
of the photon ``gravitomagnetic moment'' ({\it i.e.}, photon spin
${\bf S }$)\cite{Shen2} to the ``gravitomagnetic field'' as
follows
\begin{equation}
H=\frac{{\bf{k}}\times \dot{\bf{k}}}{k^{2}}\cdot {\bf{S}},
\eqnum{A5} \label{A5}
\end{equation}
which is just the expression (\ref{eq220}).
\\

{\bf Method iii} \quad   By using the Liouville-Von Neumann
equation

If a photon is moving inside a noncoplanarly curved optical fiber
that is wound smoothly on a large enough diameter\cite{Tomita},
then its helicity reversal does not easily take place\cite{Guo}
and the photon helicity $\frac{\bf k}{k}\cdot{\bf S}$ is therefore
conserved\cite{Chiao} and can thus be considered a
Lewis-Riesenfeld invariant $I(t)$\cite{Lewis}, which agrees with
the Liouville-Von Neumann equation (\ref{eq5}). With the help of
the spin operator commuting relations ${\bf S}\times {\bf S}=i{\bf
S}$, one can solve the Liouville-Von Neumann equation (\ref{eq5}),
namely, if the effective Hamiltonian is written as $H(t)={\bf
h}\cdot{\bf S}$, then according to the Liouville-Von Neumann
equation, one can arrive at
$\left[I(t),H(t)\right]=\left(\frac{\bf k}{k}\times{\bf
h}\right)\cdot i{\bf S}$, and readily obtain the expression for
the effective Hamiltonian (\ref{eq220}) of photons in the curved
optical fiber, {\it i.e.}, the coefficients of the effective
Hamiltonian is ${\bf h}=\frac{{\bf{k}}\times
\dot{\bf{k}}}{k^{2}}$.
\\ \\

{\bf Appendix II}. The invariant-related unitary transformation
formulation
\\

In this appendix we briefly review the invariant-related unitary
transformation formulation\cite{Gao,Gao2}, which is important in
investigating the nonadiabatic evolution process of time-dependent
quantum systems (in Sec.IV). It has been shown in Lewis and
Riesenfeld's work\cite{Lewis} that the particular solution to the
time-dependent Schr\"{o}dinger equation, the Hamiltonian
generators of which form a certain Lie algebra, is different from
the eigenstate $|\lambda, t\rangle$ of the invariant $I(t)$ (which
satisfies the Liouville-Von Neumann equation (\ref{eq5})) only by
a time-dependent $c$-number factor
$\exp[\frac{1}{i}\phi_{\lambda}(t)]$, where the time-dependent
phase is\cite{Lewis}
\begin{equation}
\phi_{\lambda}(t)=\int_{0}^{t}\langle\lambda,
t'|\left[H(t')-i\frac{\partial}{\partial t'}\right]|\lambda,
t'\rangle {\rm d}t'.
 \eqnum{A6} \label{A6}
\end{equation}
It follows that we can obtain the solution to the time-dependent
Schr\"{o}dinger equation via solving the eigenstates of the
eigenvalue equation $I(t)|\lambda, t\rangle=\lambda|\lambda,
t\rangle$. But it is difficult to solve the eigenvalue equation of
invariant $I(t)$ immediately, for the invariant is often the
linear combination of certain Lie algebraic generators and,
moreover, the coefficient factors in $I(t)$ are {\it
time-dependent}. If we can find a unitary transformation $V(t)$
that transforms the eigenvalue equation $I(t)|\lambda,
t\rangle=\lambda|\lambda, t\rangle$ into as follows
\begin{equation}
[V^{\dagger}(t)I(t)V(t)]V^{\dagger}(t)|\lambda, t\rangle=\lambda
V^{\dagger}(t)|\lambda, t\rangle,
 \eqnum{A7} \label{A7}
\end{equation}
where $V^{\dagger}(t)I(t)V(t)$ is {\it time-independent} (see, for
example, Eq.(\ref{eq9})), then the eigenstate of Eq.(\ref{A7})
corresponding to the eigenvalue $\lambda$ can be easily obtained,
which, say $|\lambda\rangle$, is also {\it time-independent},
namely, we have $|\lambda\rangle=V^{\dagger}(t)|\lambda, t\rangle$
or $|\lambda, t\rangle=V(t)|\lambda\rangle$. Correspondingly, the
time-dependent phase (\ref{A7}) is rewritten
\begin{equation}
\phi_{\lambda}(t)=\int_{0}^{t}\langle\lambda|\left[V^{\dagger}(t')H(t')V(t')-iV^{\dagger}(t')\frac{\partial}{\partial
t'}V(t')\right]|\lambda\rangle {\rm d}t'.
 \eqnum{A8} \label{A8}
\end{equation}
Thus, by making use of the above unitary transformation method, we
can obtain the exact solution to the time-dependent
Schr\"{o}dinger equation of quantum systems, the Hamiltonians of
which possess some certain Lie algebraic structures.
\\ \\

{\bf Appendix III}. The Baker-Campbell-Hausdorff formula
\\

The Baker-Campbell-Hausdorff formula is given as follows
\begin{equation}
V^{\dagger }(t)\frac{\partial }{\partial t}V(t)=\frac{\partial }{\partial t}%
L+\frac{1}{2!}[\frac{\partial }{\partial
t}L,L]+\frac{1}{3!}[[\frac{\partial
}{\partial t}L,L],L]+\frac{1}{4!}[[[\frac{\partial }{\partial t}%
L,L],L],L]+\cdots
 \eqnum{A9} \label{A9}
\end{equation}
with $V(t)=\exp [L(t)]$, $V^{\dagger }(t)=\exp [-L(t)]$, which is
of great importance for the calculation of nonadiabatic noncyclic
geometric phases of time-dependent systems.
\\


\begin{references}
\bibitem{Smith} D.R. Smith, W.J. Padilla, D.C. Vier{\it et
al.}, Phys. Rev. Lett. \textbf{84}, 4184 (2000).

\bibitem{Klimov} V.V. Klimov, Opt. Comm. \textbf{211}, 183
(2002).

\bibitem{Shelby} R.A. Shelby, D.R. Smith, and S. Schultz,
Science \textbf{292}, 77 (2001).

\bibitem{Ziolkowski2} R.W. Ziolkowski, Phys. Rev. E \textbf{64},
056625 (2001).

\bibitem{Kong} J.A. Kong, B.L. Wu, and Y. Zhang, Appl. Phys. Lett.
\textbf{80}, 2084 (2002).

\bibitem{Garcia} N. Garcia and M. Nieto-Vesperinas, Opt. Lett. \textbf{27}, 885 (2002).

\bibitem{Jianqi} J.Q. Shen, Phys. Scr. \textbf{68}, 87 (2003).

\bibitem{Veselago} V.G. Veselago, Sov. Phys. Usp. \textbf{10}, 509
(1968).

\bibitem{Gerardin} J. Gerardin and A. Lakhtakia, Phys. Lett. A \textbf{301}, 377 (2002).

\bibitem{Pendry2} J.B. Pendry, A.J. Holden, D.J. Robbins, and
W.J. Stewart, J. Phys. Condens. Matter \textbf{10}, 4785 (1998).

\bibitem{Pendry1} J.B. Pendry, A.J. Holden, W.J. Stewart, and
I. Youngs, Phys. Rev. Lett. \textbf{76}, 4773 (1996).

\bibitem{Pendry3} J.B. Pendry, A.J. Holden, D.J. Robbins, and
W.J. Stewart, IEEE Trans. Microwave Theory Tech. \textbf{ 47},
2075 (1999).

\bibitem{Maslovski} S.I. Maslovski, S.A. Tretyakov, and P.A.
Belov, Inc. Microwave Opt. Tech. Lett. \textbf{35}, 47 (2001).

\bibitem{Pendry2000} J.B. Pendry, Phys. Rev. Lett. \textbf{85}, 3966 (2000).

\bibitem{Hooft}   G.W. t' Hooft, Phys. Rev. Lett. \textbf{87}, 249701 (2001).

\bibitem{Shelby2} R.A. Shelby, D.R. Smith, S.C. Nemat-Nasser, and S. Schultz, Appl. Phys. Lett. \textbf{78}, 489 (2001).

\bibitem{Hu} L.B. Hu and S.T. Chui, Phys. Rev. B \textbf{66}, 085108 (2002).

\bibitem{Bjorken} J.D. Bjorken and S.D. Drell, {\it Relativistic Quantum
Fields} (Mc Graw-Hill Company, New York, 1965) Chap. 14.

\bibitem{Shene-print} J.Q. Shen, arXiv: cond-mat/0305414 (2003).

\bibitem{PU} J.P. Unsbo, {\it Phase Conjugation and Four-Wave
Mixing} (PhD thesis, Dept. Phys., Roy. Ins. Tech., Stockholm,
1995), Chap. 4.

\bibitem{Yablonovitch} E. Yablonovitch, Phys. Rev. Lett.
\textbf{58}, 2059 (1987).

\bibitem{Li} Z.Y. Li and Y. Xia, Phys. Rev. Lett.
\textbf{64}, 153108 (2001).

\bibitem{Berry}  M.V. Berry, Proc. Roy. Soc. London, Ser. A
\textbf{392}, 45 (1984).

\bibitem{AA}  Y. Aharonov and J. Anandan, Phys. Rev. Lett. \textbf{58}, 1593 (1987).

\bibitem{Simon}  B. Simon, Phys. Rev. Lett. \textbf{51}, 2167 (1983).

\bibitem{Furtado}  C. Furtado, V.B. Bezerra, Phys. Rev. D
\textbf{
62}, 045003 (2000).

\bibitem{Shen2}  J.Q. Shen, H.Y. Zhu, S.L. Shi, and J. Li, Phys. Scr.
\textbf{ 65}, 465 (2002).

\bibitem{Kuppermann}  Y.S. Wu, A. Kuppermann, Chem. Phys. Lett. \textbf{
201}, 178 (1993).

\bibitem{Kuppermann2}  Y.S. Wu, A. Kuppermann, Chem. Phys. Lett. \textbf{
186}, 319 (1991).

\bibitem{Levi}  B.G. Levi, Phys. Today (March), 17 (1993).

\bibitem{Wagh}  A.G. Wagh {\it et al.}, Phys. Lett. A \textbf{
268}, 209 (2000).

\bibitem{Gong}  L.F. Gong, Q. Li, and Y.L. Chen, Phys. Lett. A \textbf{
251}, 387 (1999).

\bibitem{Taguchi}  Y. Taguchi {\it et al.}, Science \textbf{
291}, 2573 (2001).

\bibitem{Falci}  G. Falci {\it et al.}, Nature \textbf{
407}, 355 (2000).

\bibitem{Wangzd} S.L. Zhu and Z.D. Wang, Phys. Rev. Lett. \textbf{89}, 097902 (2002).

\bibitem{Wangxb} X.B. Wang and M. Keiji, Phys. Rev. Lett. \textbf{87}, 097901 (2001).

\bibitem{Chiao}  R.Y. Chiao and Y.S. Wu, Phys. Rev. Lett.
\textbf{57}, 933 (1986).

\bibitem{Tomita}  A. Tomita and R.Y. Chiao, Phys. Rev. Lett. \textbf{57},
937 (1986).

\bibitem{Kwiat}  P.G. Kwiat and R.Y. Chiao, Phys. Rev. Lett. \textbf{66},
588 (1991).

\bibitem{Ross}  J.N. Ross, Opt. Quant. Elec. \textbf{16}, 455
(1984).

\bibitem{Robinson}   A.L. Robinson, Science \textbf{234}, 424 (1986).

\bibitem{Dresden}   M. Dresden and C.N. Yang, Phys. Rev. D \textbf{20},
1846 (1979).

\bibitem{Haldane1}  F.D.M. Haldane, Opt. Lett. \textbf{11}, 730
(1986).

\bibitem{Haldane2} F.D.M. Haldane, Phys. Rev. Lett. \textbf{15},
1788 (1987).

\bibitem{Chiao2}  R.Y. Chiao and A. Tomita, Phys. Rev.
Lett. \textbf{15}, 1789 (1987).

\bibitem{Zhou}  Y. Zhou, Z.H. Wu, and M. L.Ge, Chin. Phys. Lett., \textbf{16}, 316
(1999).

\bibitem{Lewis}  H.R. Lewis and W.B. Riesenfeld, J. Math. Phys. \textbf{10}, 1458 (1969).

\bibitem{Gao}  X.C. Gao, J.B. Xu, and T.Z. Qian, Phys. Rev. A \textbf{44}, 7016 (1991).

\bibitem{Gao2}  X.C. Gao, J. Fu, X.H. Li, and J. Gao, Phys. Rev. A \textbf{57}, 753
(1998).

\bibitem{Shen1} J.Q. Shen and L.H. Ma, Phys. Lett. A \textbf{308},
355 (2003).

\bibitem{Lurie} D. Luri\'{e}, {\it Particles and Fields} (Wiley, New
York, 1968) Chap. 3.

\bibitem{Shen3} J.Q. Shen, H.Y. Zhu, and H. Mao, J. Phys. Soc.
Jpn. \textbf{71}, 1440 (2002).

\bibitem{Zhu} J.Q. Shen and H.Y. Zhu, Ann. Phys.(Leipzig) \textbf{12}, 131 (2003).

\bibitem{Gaoxc} X.C. Gao, Chin. Phys. Lett. \textbf{19}, 613
(2002).

\bibitem{Bjorken2} J.D. Bjorken and S.D. Drell, {\it Relativistic Quantum
Fields} (Mc Graw-Hill Company, New York, 1965) Chap. 15.

\bibitem{Wei}  J. Wei and E. Norman, J. Math. Phys.(N.Y) \textbf{4},
575 (1963).

\bibitem{Guo} K.H. Guo and X.D. Jiang, High Ener. Phys. Nucl.
Phys.(China) \textbf{26}, 543 (2002).

\bibitem{Muller} A. Muller, H. Zbinden and N. Gisin, Euro. Phys.
Lett. \textbf{33}, 335 (1996).

\bibitem{Jones} J.A. Jones, V. Vedral, A. Ekert and G. Castagnoli, Nature \textbf{403}, 869 (2000).

\bibitem{Mashhoon}  B. Mashhoon, Phys. Lett. A \textbf{173}, 347 (1993).
\end{references}
\end{document}